\documentclass[
reprint,
superscriptaddress,
longbibliography,
 amsmath,
 amssymb,
 aps, 
 prx,
]{revtex4-2}

\usepackage{graphicx}
\usepackage{dcolumn}
\usepackage{bm}
\usepackage{amssymb}   
\usepackage{amsmath}

\usepackage{color}
\usepackage[T1]{fontenc}
\usepackage{tgbonum}
\newcommand{\be}{\begin{equation}}
\newcommand{\ee}{\end{equation}}
 
\usepackage{times}
\usepackage{soul}
\usepackage[normalem]{ulem} 
\usepackage{xcolor}

\usepackage{hyperref}
\definecolor{darkgreen}{rgb}{0.0,0.6,0.0}

\begin{document}
\title{Cell--cell Adhesion as a Double-Edged Sword in Tissue Fluidity}

\author{Anh Q. Nguyen}
\affiliation{Department of Physics, Northeastern University, Boston, MA 02115, USA}
\affiliation{Center for Theoretical Biological Physics, Northeastern University, Boston, MA 02115, USA}

\author{Pradip K. Bera}
\affiliation{Department of Mechanical Engineering, University of Wisconsin--Madison, WI, USA}

\author{Jacob Notbohm}
\affiliation{Department of Mechanical Engineering, University of Wisconsin--Madison, WI, USA}

\author{Dapeng Bi}
\email{d.bi@northeastern.edu}
\affiliation{Department of Physics, Northeastern University, Boston, MA 02115, USA}
\affiliation{Center for Theoretical Biological Physics, Northeastern University, Boston, MA 02115, USA}

\begin{abstract}
Cell migration plays a fundamental role in numerous physiological processes, including embryonic development, wound healing, and cancer metastasis. While cell--cell adhesion is known to regulate motion by shaping cell morphology and intercellular force balance, its dynamic, rate-dependent contributions to tissue behavior remain poorly understood. In this study, we examine how the dissipative nature of cell--cell adhesion influences tissue dynamics and collective migration using an extended vertex model with explicit junctional viscosity. Our findings reveal a nontrivial interplay between two distinct components of adhesion: an interfacial adhesion energy (energetic, rate-independent) contribution, which sets the effective junctional tension, and a dissipative (rate-dependent) contribution, which controls resistance to relative motion during cell rearrangements. We show that increasing the energetic component promotes migration by modifying cell shape and lowering the barrier to neighbor exchanges, whereas strengthening the dissipative component induces jamming and suppresses cell motion. Linear rheological analysis further demonstrates that, in the unjammed regime, vertex-model tissues exhibit power-law viscoelastic behavior, with adhesion modulating the power-law exponent and thereby controlling the spread of relaxation timescales. Together, these findings clarify the dual role of adhesion in governing tissue mechanics and rheology and provide a mechanistic framework for understanding the balance between fluidity and rigidity in epithelial monolayers.
\end{abstract}

\maketitle

\newpage

\section{Introduction}
\begin{figure*}[htb]
    \centering
    \includegraphics[width=1\linewidth]{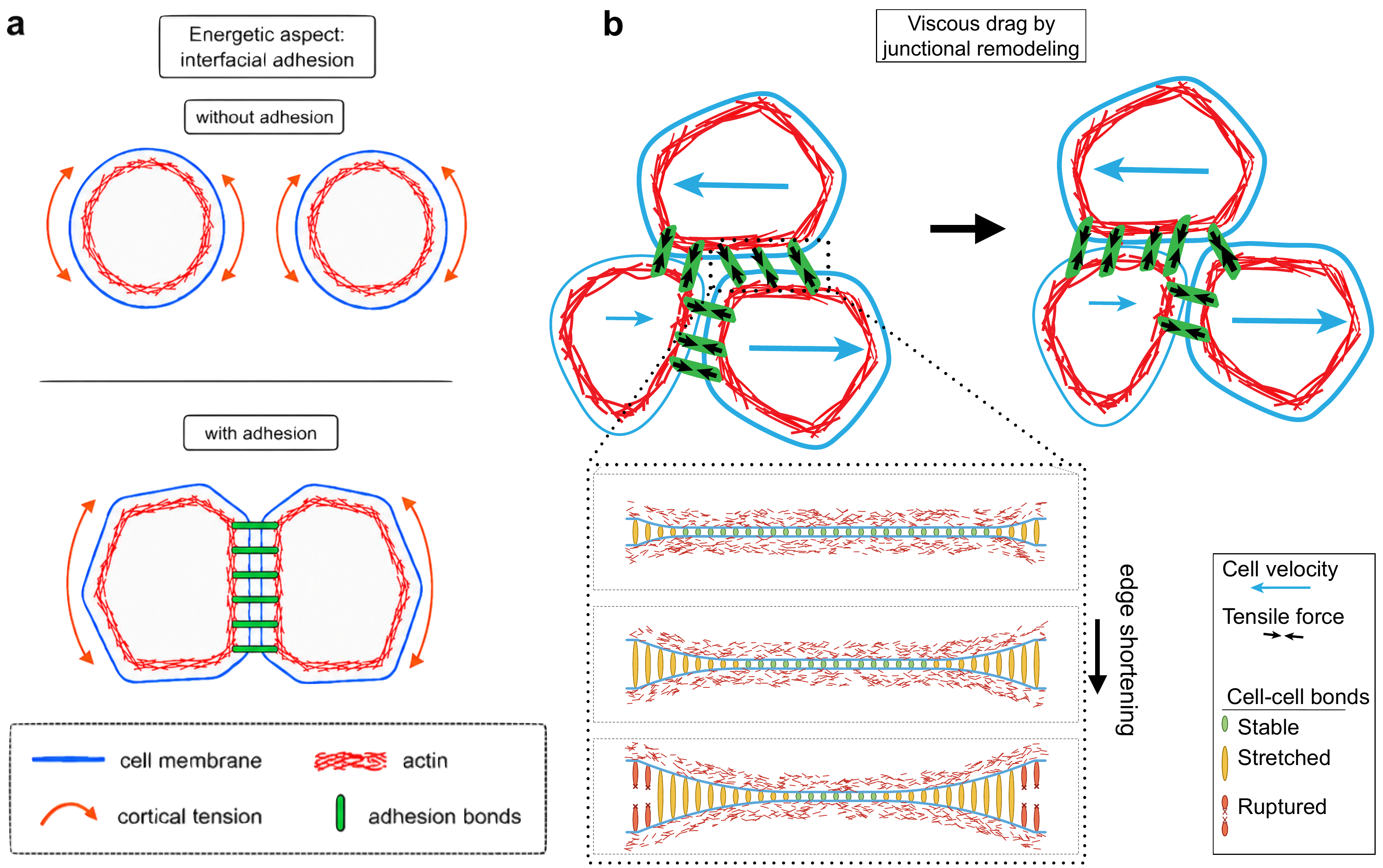}
    \caption{{Schematic illustration of the energetic and dissipative mechanical roles of cell--cell adhesion. (a) Energetic contribution of interfacial adhesion. In the absence of adhesion, contractile tension generated by the actomyosin cortex acts along the cell boundary and drives cells toward a more rounded morphology by minimizing their perimeter. This cortical contractility disfavors the formation of long, stable cell--cell interfaces. When adhesion is present, cadherin-mediated bonds form across the interface between neighboring cells and stabilize the shared junction. These adhesive interactions counteract cortical contractility by contributing an effective negative term to the junctional line tension, thereby reducing the net contractile stress at the cell--cell contact. As a result, the intercellular junction becomes elongated, the contacting cell shapes become less rounded, and the energetic barrier to local rearrangement is reduced. (b) Dissipative contribution of cell--cell adhesion. When neighboring cells move relative to one another in a confluent layer, the length of their shared contact can change. This effect is especially relevant during neighbor exchange, where one cell edge shrinks while a new edge grows. Such changes in contact length require adhesion bonds along the interface to stretch, break, and reform. The binding and unbinding of these bonds dissipates mechanical energy, producing a friction-like resistance that opposes changes in edge length.}}
    \label{adhesion mechanism}
\end{figure*}

Cell migration and tissue dynamics underpin a wide range of biological processes, from embryonic development and morphogenesis~\cite{locascio2001cell,keller2005_gastrulation,vasilyev2009collective_morp,aman2010cell,olson2018_development} to tissue repair~\cite{shaw2016wound,tetley2019tissue,aragona2017defining,vishwakarma2020mechanobiology} and cancer metastasis~\cite{yamaguchi2005cell,friedl2009collective,cheung2025collective}. A key regulator of tissue dynamics is cell--cell adhesion, which not only maintains tissue mechanical integrity~\cite{integrity_singh2009adhesion} but also mediates force transmission across cells~\cite{force_transmission_demali2014force,schoenit2025force}. The mechanical contribution of cell--cell adhesion can be understood through two complementary aspects: the interfacial adhesion energy (energetic) and the viscous-dissipative contributions. 
From the interfacial energy perspective, intercellular adhesion contributes a negative term to the effective junctional tension, thereby reducing net contractile stress along cell--cell interfaces and increasing the cell shape index~\cite{foty2005_energetic_adhesion, lecuit2007cell, manning2010coaction, winklbauer2015_energetic_adhesion}. Because higher cortical tension renders tissues more solid-like and suppresses cellular rearrangements~\cite{Perez_PRXLife.1.013004}, the adhesion-induced reduction in tension effectively promotes tissue fluidity by softening cells and lowering the energy barrier for rearrangement {(Fig.~\ref{adhesion mechanism}a). From the viscous-dissipative perspective, cell--cell adhesion dissipates energy through irreversible processes associated with the breaking and rebinding of adhesion bonds at cell--cell interfaces during relative motion and junctional remodeling (Fig.~\ref{adhesion mechanism}b). This turnover introduces resistance to relative cell motion, acting as a source of mechanical dissipation and making intercellular interactions more viscous-like~\cite{blanch2017_viscous_adhesion,fu2024regulation,arora2025viscous_important}}.

In this context, cell--cell adhesion is often metaphorically described as a “biological glue” and is widely believed to promote tissue solidity. This view has motivated the development of jamming phase diagrams inspired by adhesive colloidal particles~\cite{trappe2001jamming}, where increasing adhesion drives the system toward a more jammed, solid-like state~\cite{sadati2013collective}.  Consistent with this view, an \textit{in vitro} study has shown that the migration of human bronchial epithelial cells (HBECs) slows down as cell--cell junctions mature, implying increased adhesive strength~\cite{garcia2015}, supporting the viscous-dissipative picture. However, conflicting observations have emerged. For instance, unjamming of 4T1 epithelial monolayers under vertical compression was accompanied by increased cell--cell adhesion, as indicated by elevated cadherin density at junctions~\cite{cai2022compressive}. This finding is consistent with predictions from the conventional vertex model (VM), which captures the energetic aspect of cell--cell adhesion by modeling it as a rate-independent force that contributes to junctional tension~\cite{farhadifar2007_vm}. 

In this framework, adhesion regulates cell shape and promotes neighbor exchanges by modulating intercellular mechanical tension~\cite{bi2015density, bi2016motility}. Consequently, increased adhesion can facilitate cell rearrangements and drive tissue unjamming. Notably, this interplay between cell shape and unjamming predicted by the vertex model has been observed experimentally \textit{in vitro}, particularly in studies of asthmatic airway epithelium where increased cell elongation and rearrangement correlate with unjamming transitions~\cite{park2015unjamming}. Another example supporting this picture comes from studies of RAB5A. Elevated levels of this protein can unjam initially kinetically arrested confluent epithelial monolayers~\cite{motility_reawaken_unjamming} and fluidize epithelial spheroids~\cite{rab5a_fluidize_spheroid}. Notably, the RAB5A induced unjamming transition in confluent monolayers was accompanied by increased junctional tension, cell--cell contacts, monolayer rigidity, and E-cadherin expression~\cite{motility_reawaken_unjamming}. The effect of RAB5A is mediated through its ability to accelerate endocytosis, which increases the endo/exocytic trafficking of cadherins from the plasma membrane, effectively enhancing E-cadherin turnover and cell adhesion~\cite{ecad_regulation_endocyto,motility_reawaken_unjamming}. This outcome partly agrees with vertex model predictions, since increasing adhesion is expected to promote cell--cell contacts and unjamming. However, the vertex model would predict a reduction in junctional tension as adhesion rises, whereas the RAB5A experiments report the opposite. This discrepancy highlights a limitation of the conventional vertex framework, suggesting that adhesion dynamics and turnover may influence tissue mechanics in ways that go beyond a simple energetic contribution to junctional tension. 

{A classical framework for understanding the energetic contribution of cell--cell adhesion is the differential adhesion hypothesis (DAH)~\cite{steinberg1962mechanism,steinberg1970does}, which was later generalized into an interfacial-tension framework for tissue self-organization~\cite{brodland2002differential}. In this view, differences in effective interfacial tension, arising from the combined effects of cell--cell adhesion and cortical contractility, can drive cell sorting and tissue morphogenesis~\cite{maitre2012adhesion}. Related ideas have also been explored in preliminary vertex-model studies, where differences in interfacial tension due to cell--cell adhesion can regulate cell sorting~\cite{li2023cell_sorting}. Continuum models have similarly represented cell--cell adhesion through effective adhesive forces or interfacial-tension-like parameters, providing a complementary coarse-grained description of adhesion-driven sorting and tissue organization~\cite{armstrong2006continuum}.} 

{While the rate-independent, energetic contribution of cell--cell adhesion has been extensively studied, its rate-dependent, viscous-dissipative contribution remains comparatively underexplored, especially within vertex modeling frameworks. By mechanically coupling the contractile cortices of neighboring cells, adhesion allows stresses to be transmitted across tissues and enables coordinated cell rearrangements~\cite{lecuit2015cadherin}. Because adhesive contacts are continuously remodeled during tissue dynamics, adhesion bonds may also contribute to mechanical dissipation through bond stretching, rupture, rebinding, and cortical slippage. This microscopic picture is supported by theoretical work on cell--substrate adhesion, which has shown that adhesion-bond dynamics can generate emergent drag forces by penalizing sliding of the cell relative to the substrate~\cite{reboux2008bond,chang1999forward}. In standard vertex models, such rate-dependent resistance is typically represented phenomenologically through overdamped substrate friction or cell-level mobility. An analogous dissipative mechanism may arise at cell--cell interfaces, where dynamic adhesive bonds between apposed cortices could dissipate energy during relative intercellular motion. This interfacial contribution is difficult to represent in conventional vertex models because a single edge usually represents the shared interface between neighboring cells, precluding explicit treatment of distinct apposed cortices, cortical slippage, bond turnover, and the associated adhesion-mediated dissipation. Subcellular models such as the Apposed-Cortex Adhesion Model (ACAM) address this limitation by representing distinct apposed cortices connected by dynamic adhesion bonds~\cite{nestor2022adhesion}, allowing adhesion-mediated drag, bond turnover, and cortical slippage to emerge from microscopic cell--cell adhesion dynamics; in ACAM, the adhesion turnover rate and bond strength regulate cell rearrangements and cortical slippage~\cite{nestor2022adhesion}. Related particle-based descriptions, such as the Deformable Particle Model, additionally resolve cell shape through finite, deformable cell boundaries~\cite{boromand2018jamming}, while other recent models show that intercellular adhesion strength modulates cell rearrangements and tissue fluidity~\cite{ray2024role}. Together, these approaches support the idea that dissipative adhesion can strongly influence tissue remodeling. However, because these models retain subcellular cortical and adhesive degrees of freedom, their computational cost limits direct application to large tissue-scale systems. This motivates a coarse-grained vertex-style description that captures the dissipative contributions of cell--cell adhesion.}

In this work, we address the apparent paradox between increased adhesion and unjamming by extending the classical vertex model to include the dissipative component of cell--cell adhesion {while preserving its minimal, coarse-grained structure. Rather than introducing explicit subcellular dynamics along cell edges, we represent adhesion-mediated dissipation through an effective interfacial drag. This allows us to investigate how dynamic adhesion affects tissue migration and mechanics at the tissue scale.} Our results reveal a nontrivial interplay between the energetic and dissipative contributions of cell--cell adhesion. Specifically, increasing the energetic component facilitates cell rearrangements and enhances collective migration, whereas enhancing the dissipative component impedes motion, leading to kinetic arrest and jamming. By explicitly separating the two complementary contributions of cell--cell adhesion, the extended vertex model provides a direct theoretical framework to dissect how adhesion regulates tissue unjamming and fluidization.

{After establishing how dissipative cell--cell adhesion affects collective migration and jamming, we next ask how the same mechanism shapes tissue-scale mechanical response. This question is motivated by experimental evidence that cadherin-mediated adhesion dynamics can regulate epithelial viscoelasticity. For example, E-cadherin turnover has been shown to control stress-dependent junctional remodeling, cell rearrangements, and the effective viscosity of epithelial tissues~\cite{iyer2019epithelial}. Tissue cohesion and intercellular adhesion have also been linked to tissue viscosity and the mechanics of cell rearrangement~\cite{david2014tissue,winklbauer2015_energetic_adhesion,finegan2019tricellular,nestor2022adhesion, mccord2026}. We therefore analyze the linear rheological response of the model under step and oscillatory shear.} In the unjammed regime, the tissue behaves as a power-law viscoelastic fluid, consistent with relaxation behavior observed in MDCK monolayers~\cite{khalilgharibi2019stress_powerlaw,bonfanti2020unified}. These results establish a mechanistic link between adhesive dynamics at the cellular level and emergent tissue rheology. Together with prior studies that have focused on the energetic aspects of adhesion~\cite{tong2022_linear_rheo,huang2022shear,hertaeg2024discontinuous,nguyen2025origin}, our work helps build a more complete picture of how adhesion regulates tissue rheology.

\section{Results}
\subsection{A microscopic model of cell--cell adhesion force}
\label{microscopic model}
\begin{figure}[htb]
    \centering
    \includegraphics[width=1\linewidth]{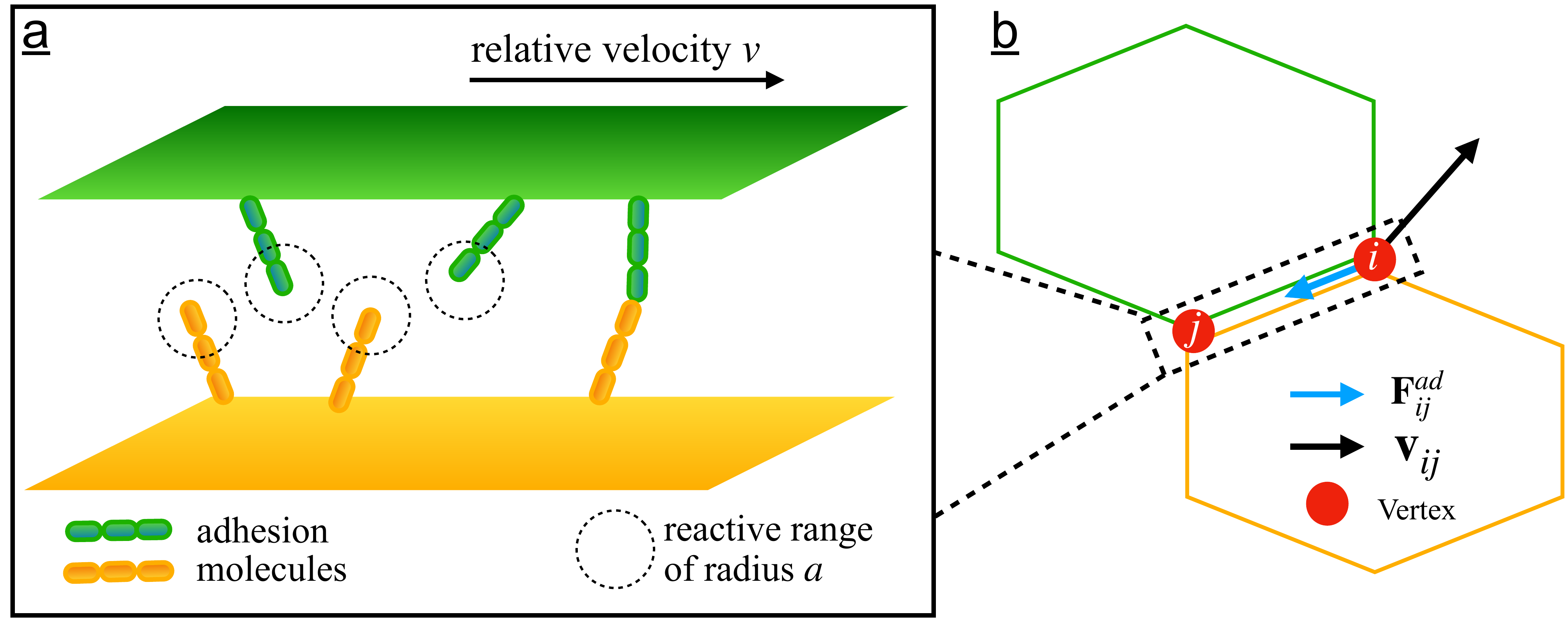}
    \caption{Modeling cell--cell adhesion force in vertex model. a) A microscopic model of cell--cell adhesion force. b) Schematic of adhesion force implementation in the vertex model.}
    \label{model_schematic}
\end{figure}

To incorporate a mechanically plausible yet computationally tractable representation of dissipative cell--cell adhesion into the vertex model, we begin from a minimal microscopic picture based on receptor–ligand binding kinetics. In this picture, cell--cell friction arises from the formation and rupture of transient adhesion bonds between receptors and ligands distributed on opposing cell membranes. In confluent epithelial tissues, the small inter-membrane separation permits direct molecular interactions across neighboring cell surfaces, enabling such transient adhesive coupling to generate dissipative forces. As two adjacent cells move relative to one another, adhesion bonds are continually broken and reformed due to the dynamic contact between their membranes (Fig.~\ref{model_schematic}a). Each time a ligand on one cell membrane encounters a receptor on the neighboring membrane, a bond may form if the interaction persists for a sufficient duration. As the relative velocity $v$ between adjacent cell surfaces increases, ligand–receptor encounters occur more frequently due to the elevated collision rate. However, the contact duration of each encounter correspondingly decreases, reducing the time available for bond formation. Under these assumptions, Chang and Hammer constructed a convection-diffusion equation that describes the dynamics of ligand-receptor interactions and derived an effective encounter rate $k_0$~\cite{chang1999forward}: 
\begin{equation}
    k_0 = 2\pi D \bigg[ \frac{I_0(\text{Pe}/2)}{K_0(\text{Pe}/2)}+2\sum_{n=1}^\infty (-1)^n\frac{I_n(\text{Pe}/2)}{K_n(\text{Pe}/2)} \bigg ]
\end{equation}
where $D$ is the diffusion coefficient of adhesion protein molecules (ligands and receptors), $I_n$ and $K_n$ are modified Bessel functions of the first and second kind respectively, and $\text{Pe}=va/D$ is the Peclet number, where $a$ is the reactive radius. {$k_0$ has units of length squared per unit time, identical to those of a diffusion coefficient.} 

Because not every encounter results in binding, the effective binding rate also depends on the probability $P$ that a receptor-ligand pair will successfully react. Assuming the receptor can occupy any position within the reactive circle ($r \leq a$) with equal likelihood, $P$ represents the chance that binding occurs before the ligand exits the interaction zone. Since the diffusion coefficient of E-cadherin in the membrane is on the order of $3\times 10^{-14}\ m^2/s$~\cite{Nelson_diffusion_coef}, while the separation between cell membranes in confluent tissues is on the order of $10\ nm$ and the typical cell speed is $1 \ \mu m/\text{min}$~\cite{trepat2009physical,angelini2011glass}, we focus on the low $\text{Pe}$ regime. In this regime, $P\sim \frac{a^2 k_{in}}{a^2k_{in}+8D}$, where $k_{in}$ is the intrinsic binding rate~\cite{chang1999forward}. Additionally, in the low $\text{Pe}$ limit, the encounter rate is well approximated by a linear function of $\text{Pe}$. The effective binding rate {of a single receptor} can be approximated by: $k_f=k_0P=\pi D m \frac{a^2 k_{in}}{a^2k_{in}+8D}\text{Pe}$, {which has dimensions of length squared per unit time,} where $m$ represents the proportionality constant between $k_0$ and $\text{Pe}$ in the low $\text{Pe}$ limit. The number of adhesion bonds $N$ between the two cell membranes evolves according to the binding–unbinding kinetics equation, $\frac{dN}{dt} = k_f \rho^2 l h - k_b N$, {where $\rho$ is the adhesion surface concentration, which has units of number of molecules per unit contact area, $l$ is the length of the contact interface, $h$ is the basal-to-apical height of the contact region, and $k_b$ is the unbinding rate, treated as a constant with dimensions of inverse time. Here, we assume equal surface concentrations of adhesion molecules on the two opposing cell membranes, so the binding rate is proportional to the product of their concentrations $\rho^2$. Assuming that each bond generates a constant force $f$, in steady state, the cell--cell frictional adhesion force can be expressed as:
\begin{equation}
    F_{fric} = fN_{steady} =\frac{\pi \rho^2 h mf}{k_b}\frac{a^3k_{in}}{a^2k_{in}+8D} lv
    \label{adhesion force}
\end{equation}
Since the force is proportional to the relative velocity, it behaves similarly to a viscous force that discourages relative motion between adjacent membranes.}

\subsection{Extended Vertex Model with Intercellular Adhesive Friction}
To understand the influence of intercellular adhesion on tissue dynamics, we study the 2D vertex model in which the biomechanical interaction is governed by the energy functional~\cite{farhadifar2007_vm,bi2015density}: $\mathcal{E}=\sum_{c=1}^N[K_A(A_c-A_0)^2+K_P(P_c-P_0)^2]$, where $N$ is the total number of cells, $K_A$ and $K_P$ represent the area and perimeter moduli, $A_c$ and $P_c$ are the area and perimeter of cell $c$, and $A_0$ and $P_0$ are the preferred area and perimeter, which we treat as homogeneous across cells.

{
To implement the dissipation by intercellular adhesion in the vertex model, we introduce a linear damping coefficient $\xi$, defined as the cell--cell adhesion drag per unit interface length per unit relative velocity, so that the microscopic result of Sec.~II.A can be written compactly as $F_{\rm fric}=\xi\,l\,v$. In the 2D vertex model, the shared interface between two adjacent cells is the edge connecting their two common vertices, $\mathbf{r}_{ij}=\mathbf{r}_i-\mathbf{r}_j$, whose length is the contact length, $l=|\mathbf{r}_{ij}|$. Although the vertex model does not resolve the two apposed cortices or their literal sliding, it does capture the quantity that sets the microscopic drag---the rate at which the contact is remodeled---through the rate of change of edge length,
\begin{equation}
v \equiv \dot{l}=\frac{d|\mathbf{r}_{ij}|}{dt}=\hat{\mathbf{r}}_{ij}\boldsymbol{\cdot}(\mathbf{v}_i-\mathbf{v}_j),
\label{edge length rate}
\end{equation}
where $\mathbf{v}_i$ is the velocity of vertex $i$ and $\hat{\mathbf{r}}_{ij}$ is the unit vector pointing from vertex $j$ to vertex $i$. Substituting $l$ and $v$ into $F_{\rm fric}=\xi\,l\,v$ gives the adhesion-mediated dissipative force exerted on the edge,
\begin{equation}
F_{ij}^{\text{ad}}=-\xi\,l\,\dot{l}\,\hat{\mathbf{r}}_{ij}=-\xi\,(\mathbf{r}_i-\mathbf{r}_j)\boldsymbol{\cdot}(\mathbf{v}_i-\mathbf{v}_j)\,\hat{\mathbf{r}}_{ij},
\label{adhesion vertex force}
\end{equation}
where the sign indicates a force that opposes changes in edge length. This coarse-graining inherits the two defining features of the microscopic drag. First, the force scales with edge length, because a longer interface contains more adhesive bonds. Secondly, it is linear in the rate of edge-length change, because that rate sets how quickly bonds are stretched, ruptured, and reformed. Because only the component of relative vertex motion along the edge enters Eq.~(\ref{adhesion vertex force}), motions that preserve edge length---rigid translations and rotations---produce no adhesive dissipation and store energy elastically, whereas junctional remodeling, in which one edge shrinks while a neighboring edge grows during a T1 transition, dissipates energy. The adhesion force thus acts as a viscous damping mechanism that resists changes in the length of the cell--cell contact interface, effectively modeling the frictional resistance arising from dynamic adhesion between adjacent cells~\cite{fu2024regulation,rozman2025vertex}.
}
The introduction of this adhesion force gives rise to the overdamped equation of motion:
\begin{equation}
\begin{aligned}
\mu\,\dot{\mathbf r}_i
&+\xi \sum_{j\in S_i}
\Bigl[
(\mathbf r_i-\mathbf r_j)\!\cdot\!(\dot{\mathbf r}_i-\dot{\mathbf r}_j)\,
\frac{\mathbf r_i-\mathbf r_j}{\lvert \mathbf r_i-\mathbf r_j\rvert}
\Bigr] \\
&= -\nabla_{\mathbf r_i}\mathcal E + \mu v_0\,\hat{\mathbf n}_i \, .
\end{aligned}
\label{EoM}
\end{equation}
where $\mu$ is the vertex-substrate viscous coefficient, $S_i$ represents the set of adjacent vertices to vertex $i$, $v_0$ is the propulsion strength and $\hat{\mathbf{n}}_i=(\cos \theta_i,\sin \theta_i)$ is the cell polarity, which  follows a random white noise rotational diffusion with mean 0 and variance $2D_r$. The model can be made dimensionless by expressing all lengths and time in units of $\sqrt {A_0}$ and $\mu/(A_0 K_A)$, respectively, resulting in a dimensionless energy functional: $e=\sum(a_c-1)^2+\kappa(p_c-p_0)^2$, where $a_c=A_c/A_0$, $p_c=P_c/\sqrt{A_0}$, $p_0=P_0/\sqrt{A_0}$, and $\kappa = K_P/(K_AA_0)$. The nondimensionalized equation of motion is:
\begin{equation}
    \mathbf{\dot r}_i + \xi_0 \sum_{j\in S_i}[(\mathbf{ r}_{ij}\boldsymbol{\cdot}\mathbf{\dot{r}}_{ij})\mathbf{\hat{ {r}}}_{ij}]=\mathbf{f}_i+ v_0\hat{n}_i
\end{equation}
where $\mathbf{ r}_{ij}=\mathbf{r}_i-\mathbf{r}_j$ represents the edge vector from vertex $j$ to vertex $i$, $\mathbf{f}_i=-\nabla_{\mathbf{r}_i}e$ is the nondimensionalized interaction force on vertex $i$, and $\xi_0=\frac{\xi \sqrt{A_0}}{\mu}$ is the dimensionless cell--cell damping coefficient. Similar phenomenological descriptions of internal viscous dissipation have recently been incorporated into vertex models to study tissue rheology, active nematics, and sustained tissue flows~\cite{tong2023_normal_mode,rozman2025vertex}. {However, in those models, viscous dissipation is effectively associated with {relative motion between vertices}, although vertices are geometric constructs rather than physical material elements. Our formulation instead localizes dissipation at cell--cell interfaces, the physical sites of adhesion complexes and the actomyosin cortex, thereby providing a physically motivated framework. In this framework, intercellular dissipation arises primarily from interfacial remodeling. Collective tissue motions or coordinated deformations that largely preserve the interfacial network generate little interfacial dissipation, whereas junction reshaping and rearrangements naturally incur a viscous cost.}
With the introduction of this adhesion force into the vertex model, the influence of cell--cell adhesion on tissue dynamics is captured through two complementary aspects: the energetic aspect, reflected in the shape index $p_0$, and the kinetic aspect, reflected in $\xi_0$.  In the simulations used for this work, the following parameters are fixed: $N=400$, $\kappa=1$, and $D_rA_0K_A/\mu = 0.5$. 

\subsection{The dual role of cell--cell adhesion in tissue dynamics}
\begin{figure}[htb]
    \centering
    \includegraphics[width=1\linewidth]{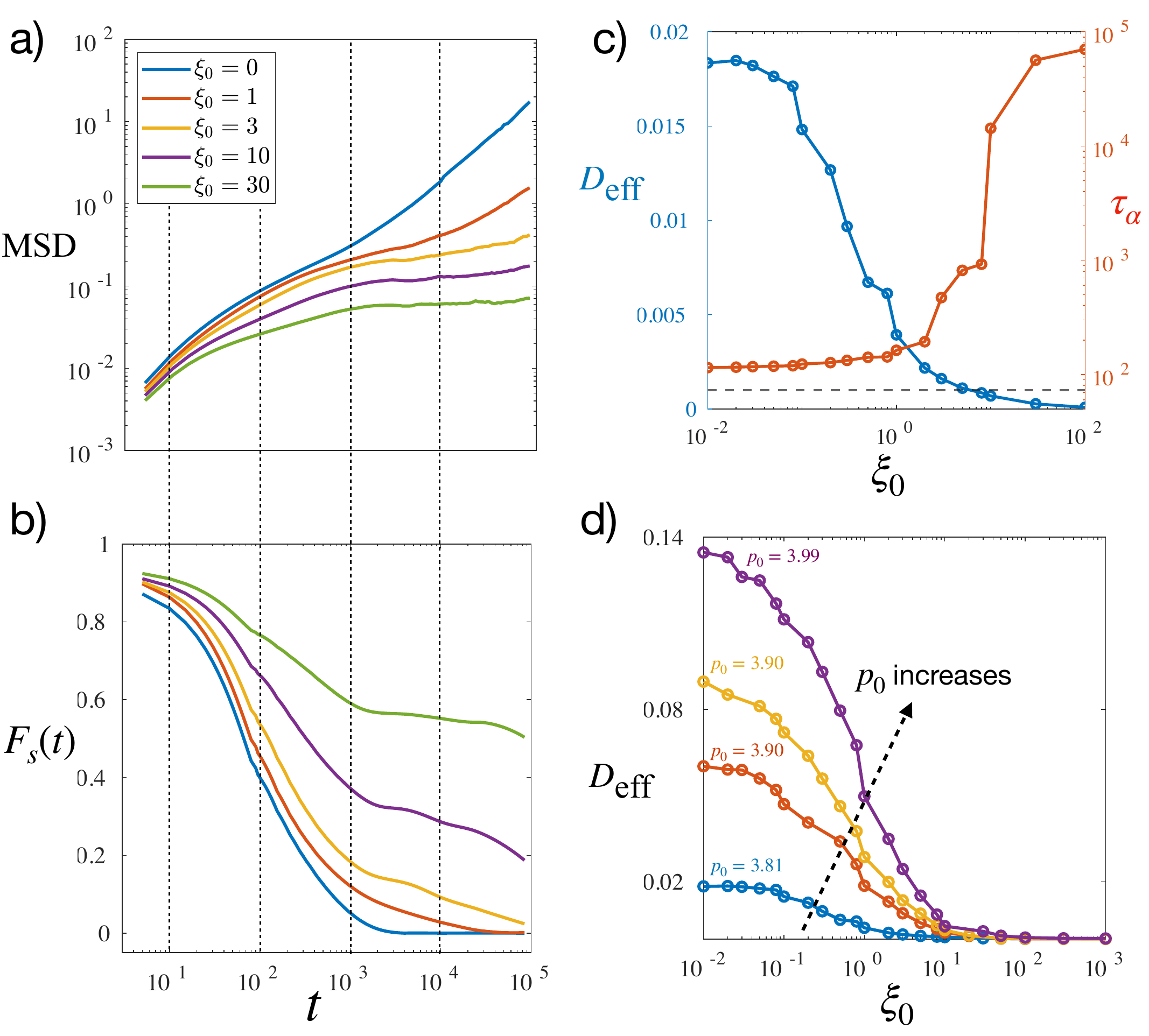}
    \caption{Glassy dynamics analysis: (a) Mean squared displacement (MSD) for $p_0 = 3.81$ and $v_0 = 0.05$, showing suppressed long-time diffusion with increasing $\xi_0$. (b) Self-intermediate scattering function $F_s(k,t)$ at the same parameters as in panel (a), evaluated at $k = \pi/\sqrt{A_0}$, illustrating the extension of the plateau and increase in relaxation time. (c) Effective diffusivity $D_{\text{eff}}$ and $\alpha$-relaxation time $\tau_\alpha$ as functions of $\xi_0$, obtained from panel (a) and (b), highlight the approach to a jammed state. (d) Dependence of $D_{\text{eff}}$ on $\xi_0$ and $p_0$ while keeping $v_0 = 0.05$, demonstrating how both energetic and viscous adhesion mechanisms influence tissue dynamics (bottom to top: $p_0=3.81, \ 3.87, \ 3.9, \ 3.99$). }
    \label{glass dynamics}
\end{figure}

\paragraph*{Adhesion-Regulated Glassy Dynamics in Simulated Epithelial Monolayers.} While the energetic aspect of cell--cell adhesion has been extensively studied in the context of tissue dynamics and jamming transitions—both theoretically~\cite{bi2015density,bi2016motility,sussman2018anomalous,huang2023bridging} and experimentally~\cite{atia2018geometric,mitchel2020primary,cai2022compressive,wang2020anisotropy}—the kinetic (viscous-dissipative) contribution remains comparatively underexplored {in the vertex modeling framework}~\cite{fu2024regulation,rozman2025vertex,ray2024role}. To systematically investigate its role, we examine the model at fixed values of $p_0 = 3.81$ and $v_0 = 0.05$. The mean squared displacement (MSD) as a function of time for varying levels of kinetic adhesion strength $\xi_0$ is shown in Fig.~\ref{glass dynamics}a. As $\xi_0$ increases, the long-time diffusive behavior of the MSD is progressively suppressed, and the intermediate-time plateau becomes more prominent and extended. This indicates stronger caging of cells by their neighbors and a shift toward glassier dynamics, where cells remain trapped in their local environment for longer times. To quantify the transition between fluid-like and jammed behavior, we use a dynamical order parameter based on the long-time self-diffusivity, defined as $D_s = \lim_{t\to\infty} \text{MSD}(t)/(4t)$. From this, the dimensionless effective diffusivity $D_{\text{eff}} = 2D_sD_r/v_0^2$~\cite{bi2016motility} is calculated. Fig.~\ref{glass dynamics}c shows how $D_{\text{eff}}$ varies with $\xi_0$. As $\xi_0$ increases, $D_{\text{eff}}$ decreases from a finite value to the numerical noise floor, signaling a transition from an unjammed, motile state to a jammed, arrested state driven by viscous cell--cell adhesion. Although the adhesion force in our model does not directly suppress cell motion, increasing $\xi_0$ effectively resists motion by penalizing changes in cell--cell junction geometry. This restricts cell rearrangement and thereby reduces large-scale tissue flow. Such adhesion-driven arrest captures the jamming behavior associated with adherens junction maturation observed in~\cite{garcia2015}.

Another standard quantity to characterize glassy dynamics is the self-intermediate scattering function~\cite{van1954correlations}, defined as $F_s(k,t)=\langle e^{i\mathbf{k}\boldsymbol{\cdot} \mathbf{\Delta r(t)}}\rangle$, where $\langle...\rangle$ represents the temporal and angular average, {and $\Delta r(t)$ is the displacement of a cell over the time interval $t$.} {Because the exponential is generally complex, the real-valued function reported in this work is obtained by taking the real part of the exponential before averaging.} Fig.~\ref{glass dynamics}b shows the relaxation of $F_s(t)$ at $k=\pi /\sqrt{A_0}$, the length scale of one cell size, representing the intrinsic cage of each cell due to its neighbors. The relaxation time $\tau_\alpha$ of $F_s$  therefore represents the timescale at which cells become uncaged and rearrange with their neighbors. Here, $\tau_\alpha$ was obtained by finding the time at which $F_s(t)$ decay to $1/e$. As $\xi_0$ increases, the system becomes more jammed, indicated by the longer relaxation time $\tau_\alpha$, with $\tau_\alpha$ eventually exceeding the simulation time, indicated by the plateau in $F_s(t)$. The dependence of $\tau_\alpha$ on $\xi_0$ is presented in Fig.~\ref{glass dynamics}c. Opposite to $D_{\text{eff}}$, $\tau_\alpha$ increases monotonically as $\xi_0$ increases, and eventually diverges as the system approaches the jamming transition. The influence of the viscous component of cell--cell adhesion, characterized by $\xi_0$, on tissue dynamics can be integrated with the well-established role of the energetic component, governed by $p_0$, which has been extensively explored in previous studies~\cite{park2015unjamming,bi2016motility,li2019mechanical,tah2025minimal}. Together, these two aspects provide a more comprehensive understanding of how cell--cell adhesion modulates tissue dynamics. As illustrated in Fig.~\ref{glass dynamics}d, increasing viscous adhesion (higher $\xi_0$) leads to enhanced jamming, consistent with stronger dissipative interactions that restrict cell rearrangements. In contrast, increasing energetic adhesion (higher $p_0$) promotes tissue fluidization by reducing junctional tension and facilitating cell neighbor exchanges~\cite{bi2015density}.

\begin{figure}[htb]
    \centering
    \includegraphics[width=1\linewidth]{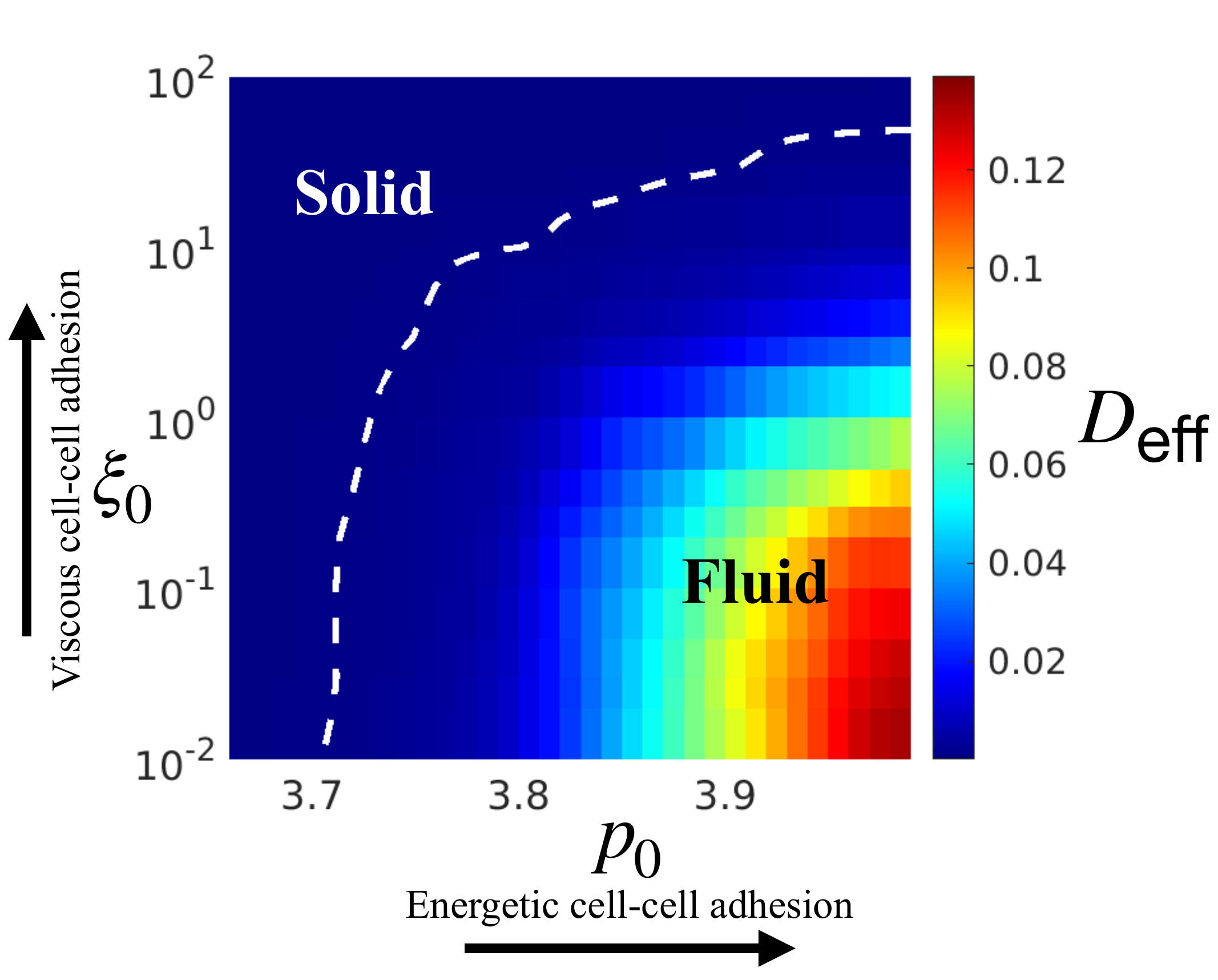}
    \caption{A jamming phase diagram governed by cell--cell adhesion. Tissue states are quantified using the effective diffusivity $D_{\text{eff}}$, with the jamming–unjamming boundary defined by the equi-diffusivity contour $D_{\text{eff}} = 0.001$ (white dashed curve). The phase diagram is constructed from simulations with $v_0=0.05$ and $D_r=0.5${.}}    
    \label{adhesion phase diagram}
\end{figure}

\paragraph*{Adhesion-Driven Jamming Phase Diagram.} The dual influence of cell--cell adhesion is summarized in the phase diagram shown in Fig.~\ref{adhesion phase diagram}, where the effective diffusivity $D_{\text{eff}}$ serves as a dynamic order parameter. We classify states with $D_{\text{eff}} \leq 10^{-3}$ as jammed and those with $D_{\text{eff}} > 10^{-3}$ as unjammed. This threshold corresponds to the random noise floor in the simulation~\cite{bi2016motility}. A corresponding jamming diagram constructed using $\tau_\alpha$ yields an essentially identical phase boundary, confirming the robustness of this classification (Fig.~\ref{tau alpha phase}). A key result from the adhesion-driven phase diagram is the non-monotonic dependence of tissue fluidity on total cell--cell adhesion: increasing energetic adhesion promotes unjamming, whereas increasing viscous adhesion has the opposite effect and drives jamming. This dual-role adhesion framework reconciles previously conflicting experimental reports on adhesion-mediated jamming~\cite{park2015unjamming,garcia2015,cai2022compressive} and underscores the importance of treating energetic and dissipative contributions as distinct, co-regulating factors in tissue mechanics.

{Although the model treats the energetic and dissipative contributions of adhesion as independently tunable parameters, these two effects are {likely} coupled in tissues because they are mediated by the same underlying adhesion machinery. Thus, increasing cell--cell adhesion in a real tissue is expected to move the system through the phase diagram along a coupled trajectory with positive slope, corresponding to simultaneous increases in both energetic and dissipative adhesion. Depending on the relative strength of these two contributions, this up-and-right trajectory may either enter a more fluid-like regime or a more solid-like regime.  How these two adhesion-mediated contributions are quantitatively coupled in real tissues remains an important open question, which we discuss in the Outlook section as a direction for future work.}

\subsection{The role of cell--cell adhesion on tissue monolayer linear rheology}
To better understand the viscoelastic property of the vertex model tissue, we  analyze the tissue linear rheology. While the effect of the energetic adhesion parameter $p_0$ on the storage and loss moduli of the vertex model has been studied previously~\cite{tong2022_linear_rheo}, the effect of the kinetic adhesion parameter $\xi_0$ on these quantities remains unexplored. To study the linear rheology of tissues, we performed strain-controlled simulations of the vertex model under oscillatory simple shear with the strain $\gamma = \gamma_0 \sin(\omega t)$. In our numerical simulations, we used $\gamma_0 = 0.001$ and set $v_0 = 0$. While recent analytical frameworks have begun to characterize the non-linear visco-elasto-plastic rheology of viscous vertex models under large deformations~\cite{Anand2026viscous}, our study deliberately focuses on the linear, small-strain regime to uncover the intrinsic spectrum of relaxation times. See \textbf{Appendix} for more details about the oscillatory deformation simulation protocol. After the system reaches a steady state, we apply a Fourier transform to the tissue shear stress $\sigma(t)$ and strain $\gamma(t)$ to calculate the complex modulus $G^*$. Details on the calculation of tissue shear stress and the complex modulus can be found in \textbf{Appendix}.

\begin{figure*}[htb]
    \centering
    \includegraphics[width=0.8\linewidth]{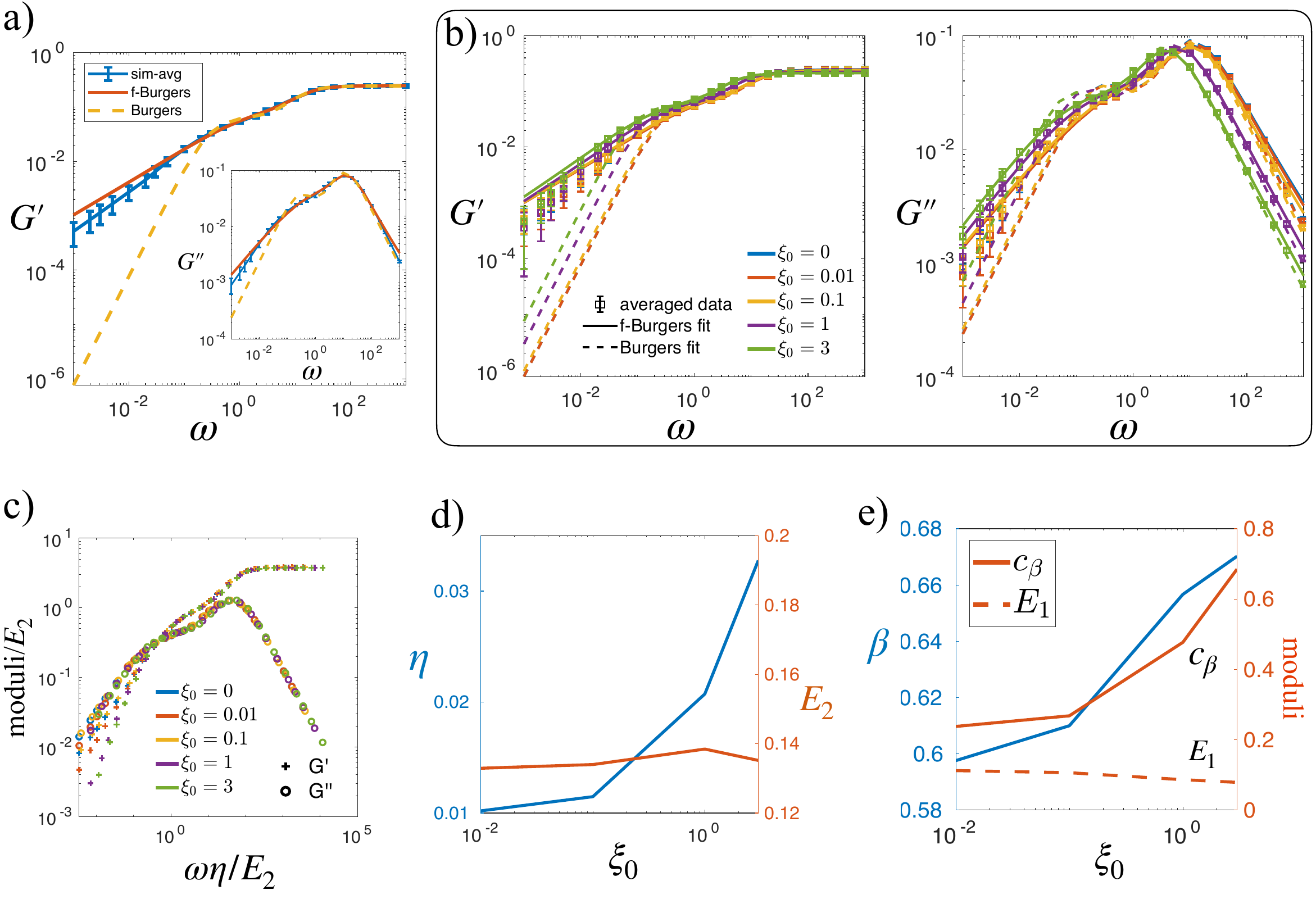}
    \caption{Linear rheology of model tissue in fluid regime at $p_0=3.9$ suggests a broad spectrum of relaxation times: 
    (a) The average complex modulus obtained from simulations at $\xi_0 = 0$ was fitted to both the classical Burgers model and the fractional Burgers model. The curve represent the mean over eight independent initial conditions, and the error bars denote the standard deviation of $G'$ and $G''$ at each frequency $\omega$. 
    (b) The complex modulus measured at different $\xi_0$. Error bars indicate the mean $\pm$ standard deviation of $G'$ and $G''$ at each $\omega$, reflecting the variability across simulations. Solid and dashed lines correspond to fits using the fractional Burgers model and the classical Burgers model, respectively.
    (c) The collapse of different moduli curves for varying cell--cell adhesion $\xi_0$. 
    (d) Viscoelastic parameters of the conventional Maxwell branch in fractional Burgers model as a function of $\xi_0$. 
    (e) Viscoelastic parameters of the fractional Maxwell branch.}
    \label{p3.9 rheology}
\end{figure*}

\paragraph*{Oscillatory Rheology of Simulated Monolayers in the Fluid Regime.} We first analyze the storage and loss moduli of the modeled tissues in the fluid-like state, with $p_0 = 3.9$ and varying $\xi_0$. Fig.~\ref{fluid complex mod} shows the complex moduli for several individual simulations at $\xi_0=0$. Interestingly, at low frequencies, the storage modulus $G'$ does not scale as $\omega^2$, and the loss modulus does not scale as $\omega$, as expected for a conventional viscoelastic fluid. Instead, both the storage and loss moduli scale as $\omega^\beta$ with $0<\beta<1$, indicating a power-law behavior that suggests a broad spectrum of relaxation times. We examined eight different simulations, and this same power-law behavior persisted across all samples. The averaged complex moduli from these simulations are shown in Fig.~\ref{p3.9 rheology}a, confirming the same scaling behavior. To highlight the difference between our modeled monolayer and a conventional viscoelastic fluid, we fit the simulated complex moduli to those predicted by the Burgers model using Eqn.~\ref{burger G star}. As presented in Fig. \ref{p3.9 rheology}(a), while the high-frequency behavior agrees with the Burgers model, clear deviations appear at low frequencies. To preserve the correct high-frequency trend while improving the fit at low frequencies, we replace one dashpot in the Burgers model with a springpot—a fractional viscoelastic element characterized by the constitutive relation $\sigma(t)=c_\beta\int_0^t(t-\tau)^{-\beta}\dot{\gamma}(\tau)d\tau$, thereby transforming the viscoelastic model into a fractional Burgers model. The analytical expression of the complex moduli of the fractional Burgers model is presented in \textbf{Appendix}. Fitting this fractional model (Eqn.~\ref{frac burgers G star}) to our simulation data yields significantly improved agreement, particularly in the low-frequency regime (Fig.~\ref{p3.9 rheology}a). Fig.~\ref{p3.9 rheology}b further illustrates that the observed power-law behavior remains robust across different cell--cell adhesion strengths $\xi_0$. 

To reveal the underlying universal viscoelastic response, we collapse the curves in {Fig.~\ref{p3.9 rheology}b} by rescaling the moduli and frequency using the fitting parameters of the conventional Maxwell branch. Specifically, the moduli are rescaled by the elastic modulus $E_2$, while the frequency is rescaled by the reciprocal of the time constant $\eta/E_2$. As shown in {Fig.~\ref{p3.9 rheology}c}, the curves collapse well in the high- and intermediate-frequency ranges, while slight deviations remain in the low-frequency regime, particularly for the storage modulus $G'$. At high frequencies, the response is dominated by the elastic contribution of the springs, resulting in an effective collapse after rescaling. In the intermediate-frequency range, the peak of $G''$ and the onset of $G'$ decay are governed by the viscous component $\eta$, which also allows the rescaled curves to align well. The residual spread in the low-frequency tails, however, suggests variations in the scaling exponent as $\xi_0$ increases. To further investigate this argument, we plot the fitting parameters of the fractional Burgers model as functions of $\xi_0$ in Fig.~\ref{p3.9 rheology}e. In this panel, the role of $\xi_0$ in capturing the viscous contribution of adhesion forces is evident from the positive correlation between $\eta$ and $\xi_0$. Moreover, $\xi_0$ also influences the springpot parameters—the exponent $\beta$ and the amplitude $c_\beta$. In contrast, while $E_1$ and $E_2$ both decrease with increasing $p_0$, as reported previously~\cite{tong2022_linear_rheo}, they remain insensitive to variations in $\xi_0$.

\begin{figure}[htb]
    \centering
    \includegraphics[width=1\linewidth]{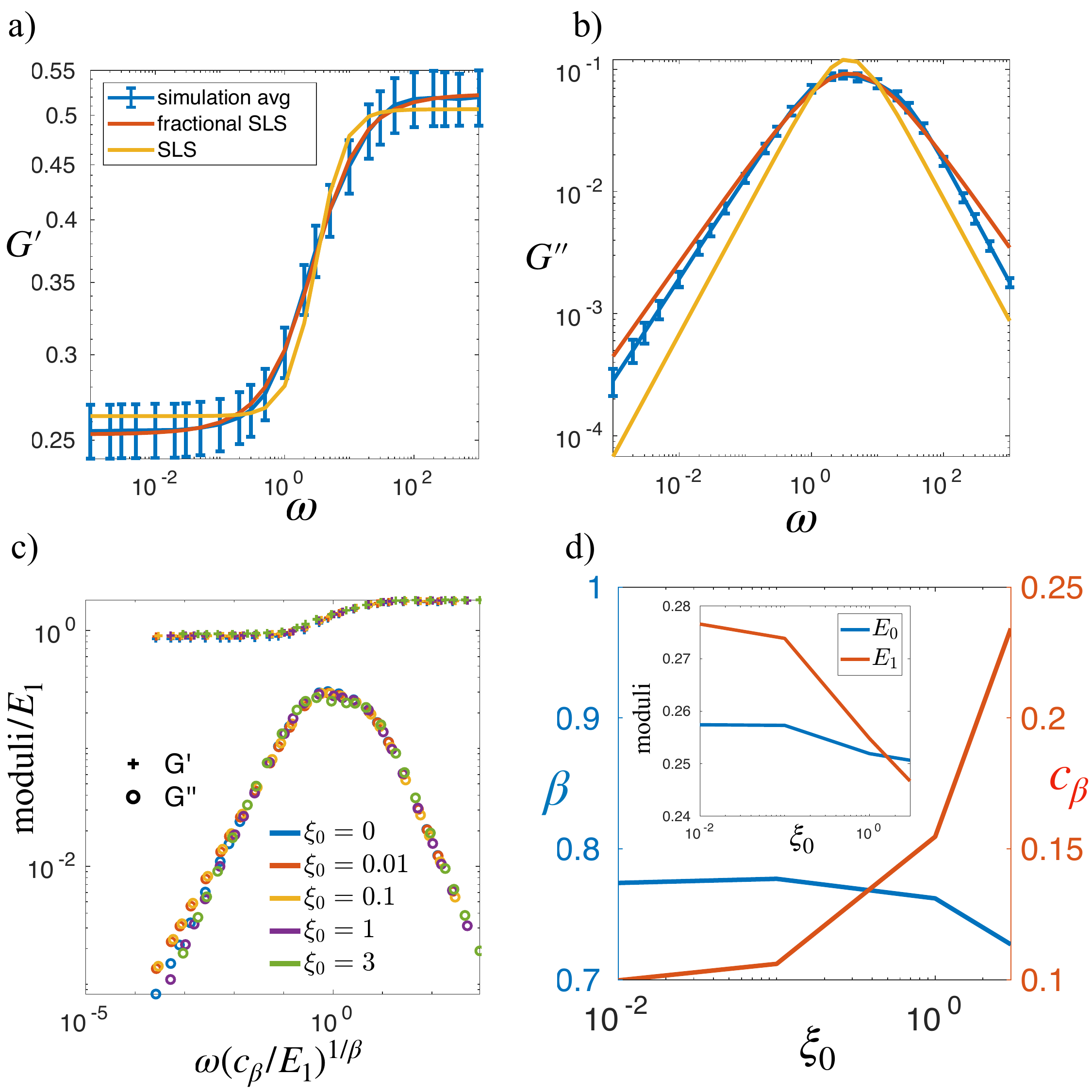}
    \caption{Linear rheology of model tissue in solid regime at $p_0=3.72$ provides another evidence for the {presence of multiple timescales in} the system.   
    (a,b) At $\xi_0=0$, the average storage ($G'$) and loss ($G''$) moduli are fitted using the fractional SLS and standard SLS models. 
    The curves in panels (a) and (b) were obtained by averaging over eight different initial conditions; error bars represent the standard deviation. 
    (c) Data collapse of different moduli curves for varying cell--cell adhesion $\xi_0$. 
    (d) Viscoelastic parameters as a function of $\xi_0$.}
    \label{p3.72 rheo}
\end{figure}

\paragraph*{Oscillatory Rheology of Simulated Monolayers in the Solid Regime.} 
To investigate the rheological response in the solid-jammed regime, we analyze the storage and loss moduli of modeled tissues at $p_0 = 3.72$ for varying $\xi_0$. Figures~\ref{p3.72 rheo}a,b show $G'(\omega)$ and $G''(\omega)$ for $\xi_0=0$. At low frequencies, the storage modulus approaches a plateau, consistent with the response of monolayers in a solid-like regime. In contrast, the loss modulus $G''$ exhibits a power-law dependence $G'' \propto \omega^\beta$ with $0<\beta<1$ instead of scaling linearly with frequency $\omega$. This fractional scaling persists across different $\xi_0$ values, as shown in Figure~\ref{p3.72 complex mod}, highlighting a robust, time-scale–rich rheological behavior. To rationalize this observation, we compare the simulation results with both the Standard Linear Solid (SLS) model and its fractional generalization, in which we replace the dashpot by a springpot. The analytical forms for the complex moduli of the SLS model and fractional SLS model are presented in Eqn.~\ref{sls G star} and Eqn.~\ref{frac sls G star}, respectively. As shown in Figure~\ref{p3.72 rheo}, the fractional SLS model captures the simulation data substantially better than the conventional SLS, particularly in reproducing the sub-linear power-law scaling of $G''$ at low frequencies. The superiority of the fractional model remains consistent across all tested $\xi_0$ values (Figure~\ref{p3.72 complex mod}). Similar sublinear scaling of $G''$ with $\beta<1$ has been reported in concentrated oil-in-water emulsions and predicted in soft-sphere models, where it has been linked to the emergence of the Boson peak in the vibrational density of states~\cite{hara2025link}. 

To evaluate the ability of the fractional SLS model to capture the rheological behavior of monolayers in the solid regime, we attempt to collapse the complex moduli at different values of $\xi_0$ onto a single master curve. Unlike the fractional Burgers model—where the dashpot provides a natural relaxation timescale—the fractional SLS model contains no intrinsic characteristic time; its frequency dependence is controlled entirely by the springpot. Consequently, the appropriate rescaling must be constructed from the parameters of the fractional Maxwell branch. We therefore rescale the moduli by the elastic modulus $E_1$ and rescale the frequency by the springpot-derived timescale $(E_1/c_\beta)^{1/\beta}$, which follows directly from the dimensional structure of the fractional term. As shown in Fig.~\ref{p3.72 rheo}c, this rescaling produces an excellent collapse of the moduli across all tested $\xi_0$ values, in both the low- and high-frequency regimes, demonstrating that the response is governed by a common underlying scaling form. Notably, in contrast to the deviation between different rescaled $G''$ {at low frequency in the fluid-like regime (Fig.~\ref{p3.9 rheology}c)}, the solid-regime data collapse well {(Fig.~\ref{p3.72 rheo}c)}, suggesting that the scaling exponent $\beta$ is insensitive to $\xi_0$. This is confirmed by the fitted parameters of the fractional SLS model (Fig.~\ref{p3.72 rheo}d): the exponent $\beta$ and the elastic modulus $E_0$ vary weakly with $\xi_0$. The springpot amplitude $c_\beta$ increases while the modulus $E_1$ decreases with increasing $\xi_0$, indicating that stronger cell--cell adhesion enhances the contribution of the springpot to the overall tissue rheology. Since the viscous response of the fractional SLS model used here arises solely from the springpot, the increase of $c_\beta$ with $\xi_0$ indicates that in the solid-like regime, stronger cell--cell adhesion produces a larger effective viscosity, consistent with the role of $\xi_0$ as the parameter of junctional viscous dissipation in the vertex model. The robustness of the observed sublinear power-law scaling of the loss modulus {at low frequency, $G''(\omega)\sim\omega^\beta$ with $\beta<1$,} in both the solid and fluid regimes suggests that a broad distribution of relaxation times is an intrinsic feature of tissue monolayers within the vertex-model framework. {We hypothesize that this broad spectrum is closely related to structural disorder in the monolayer. Unlike ordered hexagonal configurations, whose rheology can be well described by the SLS and Burgers models~\cite{tong2022_linear_rheo}, the configurations studied here are disordered and contain a mixture of hexagonal and non-hexagonal cells; different local cellular environments may therefore relax on different timescales, giving rise to the effective springpot-like behavior observed in the fitted fractional models. In this interpretation, the fractional Burgers model provides a natural unified framework for the fluid regime, since it contains the classical Burgers model as the ordered-fluid limit when the fractional exponent approaches unity, while the fully fractional Burgers model captures a fluid-like material with power-law relaxation that may correspond to a disordered fluid. Similarly, in the solid regime, the fractional standard linear solid captures a disordered solid with power-law relaxation, whereas the classical standard linear solid is recovered when the springpot contribution is removed, corresponding to a more ordered solid with a narrower relaxation spectrum. Thus, the fractional exponent may encode information about the degree of structural disorder in the tissue, linking the observed power-law rheology to the heterogeneity of local cellular environments. This connection between structural disorder, relaxation spectra, and fractional rheology will be investigated systematically in future work.}

\begin{figure*}[htb]
    \centering
    \includegraphics[width=0.9\linewidth]{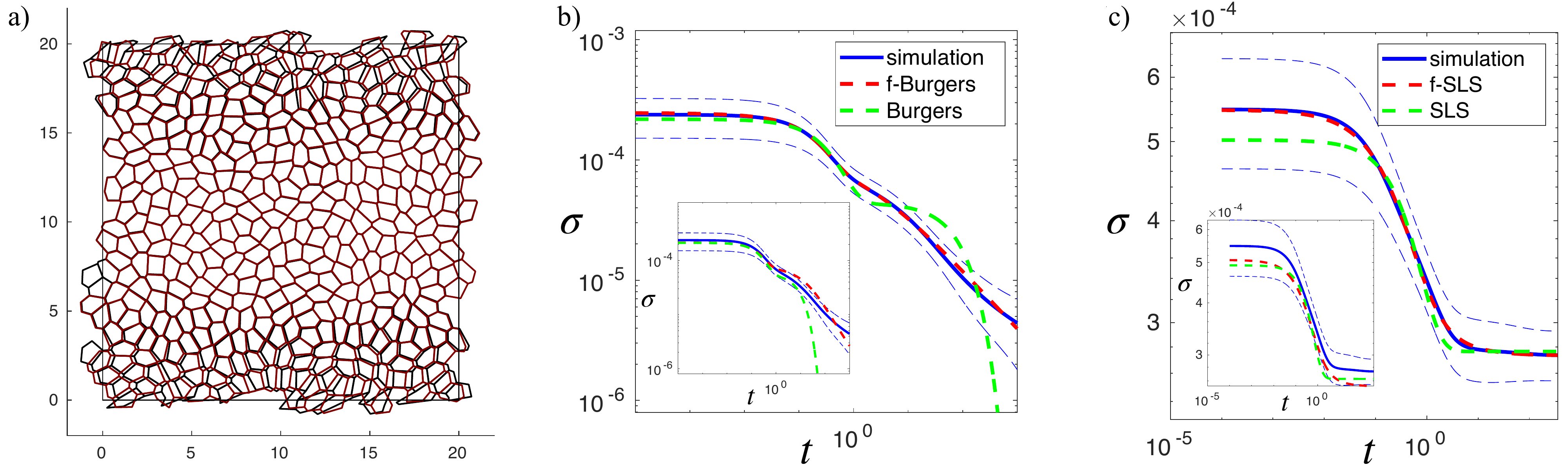}
    \caption{Relaxation modulus provides an independent confirmation of power-law rheology. a) Superimposed snapshots of the tissue configuration immediately after the application of a step strain and after long-time relaxation. Black polygons denote cell shapes directly following the imposed deformation, whereas red polygons correspond to the relaxed steady-state configuration. (b) Averaged stress relaxation after creep for monolayers at $p_0 = 3.9$ and $\xi_0=1$. Blue-thin dashed lines represent the variation in the stress between samples. The relaxation response is fitted using the fractional Burgers model and the conventional Burgers model. Inset: Relaxation moduli evaluated using parameters extracted from oscillatory {simulations}; the same legend applies. (c) Averaged stress relaxation following a step strain for monolayers at $p_0 = 3.72$ and $\xi_0=1$. Blue-thin dashed lines represent the variation in the stress between samples. The relaxation response is fitted using the fractional SLS model and the conventional SLS model. Inset: Relaxation moduli evaluated using parameters extracted from oscillatory {simulations}; the same legend applies.}
    \label{relax mod}
\end{figure*}

\paragraph*{Stress Relaxation of Simulated Monolayers Under Step Strain.} Motivated by the low-frequency scaling of the complex moduli, we examine the relaxation modulus $G(t)$ and expect a corresponding power-law decay. To test this prediction, we perform stress-relaxation simulations by imposing a small step strain of $\gamma_0 = 10^{-4}$ and holding it fixed while monitoring the resulting stress decay. To isolate the intrinsic material response, T1 transitions are prohibited during relaxation. The resulting curves, $\sigma(t)=\gamma_0 G(t)$, for monolayers with $p_0 = 3.9$ and $3.72$ at $\xi_0 = 1$, are shown in Fig.~\ref{relax mod}b and c, respectively. As anticipated, the relaxation modulus at $p_0=3.9$ displays a power-law decay, consistent with a fluid-like material lacking a single characteristic relaxation timescale. Power-law relaxation in biological systems has been reported at both the cellular scale~\cite{fischer2016rheology} and the tissue scale~\cite{khalilgharibi2019stress_powerlaw,bonfanti2020unified}. Fitting these curves to the Burgers and fractional Burgers models shows that the fractional model yields a substantially improved description (Fig.~\ref{relax mod}b), a result that holds across all tested values of $\xi_0$. Because a power-law decay in $G(t)$ implies a power-law growth in the creep compliance with the same exponent $\beta$, this observation aligns with recent experiments on \textit{Drosophila} embryonic epithelium. Using ferrofluid droplets under a constant pulling force, the displacement-versus-time relation of these droplets embedded in tissue captures the tissue's creep compliance; a scaling exponent of $\beta \approx 1/2$ has been reported~\cite{cheikh2025scaling}.

In the solid regime, the fractional SLS model likewise provides a superior fit to the simulated step-strain response relative to the conventional SLS model (Fig.~\ref{relax mod}c). The improved performance of the fractional viscoelastic models in both fluid-like and solid-like regimes further supports the interpretation that tissue monolayers possess a broad distribution of relaxation times. We also use the fitting parameters obtained from oscillatory measurements to predict the stress-relaxation response (insets of Fig.~\ref{relax mod}b,c). Although these predictions do not perfectly reproduce the relaxation curves, they remain considerably more accurate than predictions from the corresponding conventional models. The discrepancies between parameters extracted from oscillatory and relaxation protocols are unsurprising. In the oscillatory {simulations}, the continuous cyclic driving effectively mechanically anneals the tissue: repeated oscillations progressively relax internal stresses, guide the system toward a smoother energy landscape, and place it in a more homogeneous, well-annealed mechanical state. In contrast, the relaxation {simulations} begin only from an instantaneously equilibrated configuration, which is equilibrated in the sense of mechanical force balance but is not necessarily stress-free and may retain residual disorder inherited from the initial condition. Because no mechanical cycling occurs during relaxation, the system remains in this comparatively poorly annealed state throughout the step-strain test. Consequently, the material parameters inferred from relaxation data reflect these less-annealed configurations, whereas the oscillatory measurements probe more mechanically conditioned states.

\section{Discussion}
In this work, we addressed a longstanding paradox regarding the role of cell--cell adhesion in tissue dynamics, which can be characterized using either the effective diffusivity $D_{\text{eff}}$ or the uncaging relaxation time $\tau_\alpha$, both of which are experimentally accessible measures, and we examined how adhesion regulates the viscoelastic response of epithelial monolayers within the vertex-model framework. Our approach explicitly disentangles two mechanistically distinct components of adhesion: an energetic contribution, which reduces junctional tension, and a kinetic contribution, which resists relative motion across the shared cell--cell interface. The latter is incorporated into the vertex model with support from a microscopic description showing that relative motion between adjacent cell surfaces, together with adhesion-bond forces, generates an effective mesoscopic drag. We demonstrated that increasing the kinetic component of adhesion progressively enhances dynamical arrest, ultimately driving the system into a jammed state. However, this jamming effect competes with the fluidization induced by the energetic adhesion component, which has been extensively characterized~\cite{bi2015density,bi2016motility}. As a result of this competition, the tissue exhibits a non-monotonic dependence on adhesion strength, such that increasing adhesion can initially promote unjamming before leading to jamming at higher values. The corresponding phase diagram captures this complex, non-monotonic jamming–unjamming behavior and delineates the regimes in which either mechanism dominates. By capturing the competing effects of energetic and kinetic adhesion, our framework resolves previously contradictory experimental findings on adhesion-driven jamming in biological tissues~\cite{fu2024regulation,garcia2015,cai2022compressive,park2015unjamming} and provides a unified perspective on how adhesion modulates cell motion and tissue dynamics.

We further investigated how cell--cell adhesion influences the rheological properties of vertex-model monolayers subjected to both oscillatory and step-strain deformation. The resulting stress responses in each case exhibit power-law behavior, motivating their analysis using fractional viscoelastic models. The power-law behavior reported here further suggests that rich, multiscale relaxation dynamics can emerge even within a minimal framework such as the vertex model. This behavior becomes more pronounced when the response is averaged over multiple realizations. Because the model is self-averaging—so that ensemble averaging over samples is equivalent to spatial averaging over different regions of the monolayer—the enhanced prominence of the power law suggests that the vertex-model tissue is intrinsically heterogeneous in its local relaxation timescales. A systematic finite-size analysis of this heterogeneity, including the distribution and spatial organization of local relaxation times, would help identify the microscopic origin of the observed multiscale behavior and further elucidate why a broad spectrum of relaxation processes is an inherent feature of tissue monolayers within the vertex-model framework. More broadly, the linear rheological response observed in our simulations reinforces the view that epithelial tissues behave as viscoelastic materials with non-exponential, scale-free relaxation. Analysis within the extended vertex model indicates that, by modulating both elastic storage and viscous dissipation, cell--cell adhesion emerges as a key regulator of epithelial monolayer viscoelasticity and dynamical state, with the capacity to soften the tissue while simultaneously enhancing its viscous properties.

It is important to note that the absolute glassy arrest observed in our framework is inherently tied to the isotropic active driving and 2D topological constraints of the standard vertex model. For instance, recent multiphase field models demonstrate that when motility is coupled with intercellular friction, the resulting spatial flow correlations can drive macroscopic orientational order rather than simple arrest~\cite{Chiang2024}. Furthermore, 3D active foam models suggest that out-of-plane degrees of freedom and continuous cell--cell debonding may prevent the tissue from fully solidifying even under high intercellular friction~\cite{vangheel2026rigidity}. Reconciling these dimensionality and hydrodynamic effects with our dual-role adhesion framework represents an important direction for future research.

\subsection*{Quantitative Predictions for Experiments}
{To illustrate the predictive scope of the extended vertex model, we apply it to several experimentally reported phenomena. In particular, the dual-role formulation of adhesion provides a unified interpretation of the non-monotonic dependence of cell shape on cell--cell adhesion level—quantified by junctional E-cadherin intensity—in the \textit{Drosophila} embryo during axis elongation~\cite{wang2024cadherin}. In that study, one perturbation reduced adhesion and another increased it relative to wild type, yet both conditions exhibited elevated cell shape indices compared to wild type. Within our framework, this seemingly paradoxical behavior arises because adhesion modulates tissue mechanics through separable energetic and dissipative contributions. Changes in adhesion can therefore shift the system through distinct regions of the phase diagram without requiring a monotonic geometric response. A representative trajectory capturing this qualitative behavior is shown in Fig.~\ref{hypo q path}}

{We further connect the rheological predictions of the extended vertex model to experimental measurements. Specifically, we compared the relaxation timescales obtained in simulations with measurements in microtissues reported by Walker et al.~\cite{walker2020time}. In those experiments, microtissues were subjected to step-strain deformations using cantilevers and allowed to relax under fixed strain. Using the fitted parameters of the Standard Linear Solid (SLS) model in the solid regime (Fig.~\ref{sls time scale}) together with the time calibration reported in~\cite{bera2026shape}, we converted the characteristic relaxation time $\tau = \eta/E_1$ into physical units. The resulting relaxation times, which vary with $\xi_0$ and lie in the range of approximately 6–20 seconds, are comparable to the experimentally reported value of roughly 14 seconds. This level of agreement supports the physical interpretability of the model parameters and suggests that the emergent rheological timescales are within a quantitatively reasonable range.}

{Additionally, the dependence of the rheological parameters $\beta,\ c_\beta$, and $\eta$ on the cell--cell damping coefficient $\xi_0$, shown in Fig.~\ref{p3.9 rheology}, establishes a direct link between vertex-model predictions and observations in MDCK monolayers treated with the Y27632 ROCK inhibitor, a biochemical perturbation known to substantially increase cell--cell adhesion~\cite{diao2015_rho_adhesion,pipparelli2013rock_adhesion}. In these works, the relaxation modulus $G(t)$ was analyzed using a fractional viscoelastic model analogous to our fractional Burgers formulation, and Y27632 treatment produced a marked increase—approximately two-fold—in the effective viscosity $\eta$, accompanied by a reduction in the elastic modulus $E_2$~\cite{bonfanti2020unified}. While the adhesion-induced softening is consistent with predictions from the conventional vertex model~\cite{tong2022_linear_rheo}, the concomitant increase in viscosity is not captured within that framework. By contrast, the extended model naturally reproduces this effect through the explicit dependence of both the springpot amplitude $c_\beta$ and the viscosity $\eta$ on $\xi_0$. This result reinforces the conclusion that viscous dissipation at cell--cell junctions is a central determinant of epithelial monolayer rheology within the vertex-model description.} 

The examples above are preliminary demonstrations of the model’s scope rather than a systematic or exhaustive mapping of parameter space to experimental observables. A more comprehensive and quantitatively rigorous comparison with experiments is presented in a companion study~\cite{bera2026shape}{, which provides a direct test of the role of cell--cell adhesion in tissue dynamics. In those experiments, cell--cell adhesion was perturbed in MDCK monolayers while other major determinants of tissue fluidity, including cell density, traction forces, intercellular tension, and cell shape, remained effectively unchanged. Thus, the observed changes in collective dynamics can be attributed primarily to the altered cell--cell adhesion state, rather than to concurrent changes in geometry, active driving, or cell--substrate mechanics. Under these controlled conditions, perturbing adhesion produced substantial changes in tissue fluidity without a corresponding change in cell shape index. This shape-independent fluidization cannot be reconciled with the original vertex model, in which cell geometry is the primary order parameter {for tissue fluidity}. In contrast, the extended model captures this behavior by incorporating viscous adhesion as an independent mechanical contribution. The observation that tissue fluidity can be modulated independently of cell shape challenges the prevailing geometric paradigm and motivates a broader framework in which dissipative mechanisms play a central role in governing epithelial dynamics. Together with the qualitative application to adhesion-dependent morphogenetic flows in the \textit{Drosophila} embryo (Fig.~\ref{hypo q path}), these results illustrate how the dual energetic--dissipative formulation connects adhesion perturbations to tissue-scale dynamics across epithelial systems.}

\subsection*{Outlook} 
In many experimental studies of tissue jamming and unjamming~\cite{angelini2011glass,park2015unjamming,garcia2015,tetley2019tissue, saraswathibhatla2020, cai2022compressive}, cell–substrate mechanical interactions play a central role in addition to cell--cell adhesion. Traction forces, substrate stiffness, and active motility collectively influence stress generation, energy dissipation, and collective rearrangements. Within the vertex-model framework, these active effects are commonly coarse-grained into the motility parameter $v_0$, while dissipative interactions with the substrate are captured by a baseline friction coefficient ($\mu$). The present work isolates the contribution of cell--cell adhesion in order to clarify its dual energetic and dissipative roles. However, future extensions of this model could comprehensively explore the full spectrum of cell–substrate coupling. On one end, incorporating the effect of $v_0$ would naturally extend the current adhesion-controlled phase diagram into a three-dimensional jamming landscape, enabling systematic exploration of how active driving and intercellular coupling cooperate to regulate yielding and collective flow. {Preliminary simulations suggest that increasing $v_0$ enhances tissue fluidity, producing trends in $D_{\rm eff}$ versus $\xi_0$ similar to those observed when $p_0$ is increased as shown in Fig. \ref{glass dynamics}d. Consequently, larger $v_0$ is expected to shift the jamming transition toward higher dissipative adhesion, indicating that stronger cell--cell adhesion is required to arrest more actively driven tissues.} On the opposite extreme, to investigate completely free-floating epithelia, our dual-role adhesion framework could be integrated with recent theoretical advances that formulate rotationally invariant viscous vertex matrices capable of simulating zero-substrate-friction dynamics~\cite{lin2026viscous}.

An important aspect of tissue dynamics that is not addressed here is {coordinated} cell motion, which can be quantified by the velocity--velocity correlation length. Experimental studies have shown that as cell--cell adhesion increases, the correlation length initially grows but eventually decreases at longer times~\cite{garcia2015}. The initial positive correlation between collective motion and adhesion is consistent with results from vertex models that incorporate vertex--vertex viscous interactions~\cite{rozman2025vertex}. However, the subsequent decrease in correlation length observed at longer times remains poorly understood. One possible explanation is that adhesion proteins may influence not only the energetic adhesion parameter $p_0$ and the dissipative adhesion parameter $\xi_0$, but also the magnitude of stochastic fluctuations in the system, represented in our extended model by the rotational noise strength $D_r$. In this scenario, changes in adhesion could simultaneously modify the mechanical coupling between cells and the level of active fluctuations, giving rise to the observed non-monotonic behavior in collective motion. A more detailed analysis of the relationship between collective motion and jamming will be explored in future work.

{As cell--cell adhesion has both energetic and dissipative aspects, how these two contributions are coupled in real tissues is an important question that remains poorly understood. We hypothesize that different cadherin types may preferentially regulate different aspects of adhesion because they differ in bond strength, turnover rate, cortical coupling, and relaxation timescale. Consistent with this idea, partial EMT in primary human bronchial epithelial cells is associated with increased N-cadherin expression and a larger cell shape index~\cite{mitchel2020primary}. Because cell shape is closely linked to the energetic aspect of adhesion, this correlation suggests that N-cadherin may contribute primarily to the energetic component. However, this interpretation remains tentative because partial EMT also involves reduced and disrupted E-cadherin localization, which may also affect adhesion-mediated mechanics. Additional support comes from our companion work~\cite{bera2026shape}, where perturbing E-cadherin significantly changes tissue dynamics with little change in cell shape or other tissue properties, suggesting that E-cadherin primarily affects the viscous component in that system. Although this hypothesis provides a possible interpretation of energetic--dissipative adhesion coupling, testing it systematically would require independently perturbing different cadherin types while holding cell density, motility, traction forces, cortical tension, and substrate mechanics fixed. This remains experimentally challenging because cadherin perturbations can also alter junctional signaling, cytoskeletal organization, cell polarity, and force transmission.}

\begin{acknowledgments}
A.N. and D.B. acknowledge support from NSF Grants No. DMR-2046683 and PHY-2019745, the Sloan Research Fellowship, NIH Grant No. R35GM150494 and the Human Frontier Science Program research grant (RGP0007/2022). J.N. acknowledges support from NSF Grant No. CMMI-2205141 and NIH Grant No. R35GM151171. 
\end{acknowledgments}

\bibliography{references.bib}

\section*{Code availability}
The custom code developed for the simulation in this study is available in Zenodo repository~\cite{nqanh1995_2026_21536426}.

\section*{Appendix}
\setcounter{section}{0}
\renewcommand{\thesection}{}
\renewcommand{\thesubsection}{\Alph{subsection}}
\renewcommand\thefigure{S\arabic{figure}}
\renewcommand{\theHfigure}{S\arabic{figure}}
\setcounter{figure}{0}

\subsection{Solving the equation of motion}
Due to the adhesion force implemented, the $x$ and $y$ directions are coupled. The equation of motion can be rewritten in matrix form as:
\begin{equation}
    \mathbf{Mv=f}
    \label{matrix equation of motion}
\end{equation}
For a system of size $N$, $\mathbf{f}$ is a $4N\times 1$ column vector that represents the interaction force on each vertex, with the first $2N$ entries being the $x$-components of the forces and the remaining $2N$ entries being the $y$-components of the forces. Similarly, $\mathbf{v}$ is a $4N\times 1$ column vector storing the velocities of the vertices. $\mathbf{M}$ is a $4N\times 4N$ sparse matrix, which can be expressed as:
\begin{equation*}
    M=\begin{bmatrix}
    M^{xx} & M^{yx}\\
    M^{xy} & M^{yy}
\end{bmatrix}
\end{equation*}
Where $M^{\alpha\beta}$, with $\alpha,\beta\in \{x,y\}$, is a $2N\times 2N$ matrix encoding the coupling between the velocities in the $\alpha$-direction and the force in the $\beta$-direction. The non-zero elements in those sub-matrices are $M_{ii}^{\alpha \beta}=\delta_{\alpha,\beta}+\xi_0\sum_{j\in S_i}{r_{ij}^\alpha r_{ij}^\beta}/{|\mathbf{r}_{ij}|}$  and $M^{\alpha \beta}_{ij}=-\xi_0 r_{ij}^\alpha r_{ij}^\beta /{|\mathbf{r}_{ij}|}$ for $j\in S_i$. $M^{\alpha \beta}$ can therefore be conveniently constructed from the $4N\times 1$ vertex position vector $\mathbf{r}(t)$ and the vertex-vertex adjacency matrix. The equation of motion can be integrated numerically as 
\begin{equation}
    \mathbf{r}(t+\Delta t)=\mathbf{r}(t)+\Delta t\mathbf{M}^{-1}\mathbf{f}(t)
    \label{integrating eom}
\end{equation}

\subsection{The stress tensor for each cell in the presence of cell--cell adhesion force}

In our model, the stress tensor $\boldsymbol{\sigma}_c$ for cell $c$ is expressed as:
\begin{equation}
    \boldsymbol{\sigma_{c}}=\boldsymbol\sigma_c^{int }+\boldsymbol\sigma_c^{ad}
\end{equation}
where $\boldsymbol\sigma_c^{int}$ is the stress tensor due to the interaction force, which can be computed as\cite{yang2017correlating}:

\begin{equation}
    \boldsymbol\sigma_c^{int}=K_A(A_c-A_0)\mathbf{I}+\frac{1}{A_c}\sum _{e\in c }K_P(P_C-P_0)\mathbf{\hat L}_e \otimes \mathbf{L}_e
\end{equation}
where $\mathbf{I}$ is the unit tensor, $\mathbf{L}_e$ is edge vector of edge $e$, the sum runs over ${e}\in c$, the set of edges of cell $c$, and $\otimes$ is the tensor product notation. $\boldsymbol \sigma_c^{ad}$ is the stress tensor due to the adhesion force, which can be computed as\cite{tong2023_normal_mode}:
\begin{equation}
    \boldsymbol\sigma_c^{ad}=-\frac{1}{2z_cA_c} \sum_{i\in c}\big(\mathbf{R_i}\otimes \mathbf{F}^{ad}_i+\mathbf{F}_i^{ad}\otimes \mathbf{R}_i\big)
\end{equation}
where $z_c$ is the number of vertices cell $c$ has, $\mathbf{R}_i=\mathbf{r}_i-\mathbf{r}_c$ is the position of vertex $i$ relative to cell $c$, and $\mathbf{F}_i^{ad}$ is the total adhesion force at vertex $i$. 
The tissue stress tensor $\boldsymbol{\sigma}$ was then computed as:
\begin{equation}
    \boldsymbol{\sigma} = \frac{\sum_cA_c\sigma_c}{\sum_cA_c}
    \label{tissue stress tensor}
\end{equation}
The tissue shear stress was then extracted from the tissue stress tensor as $\sigma=\boldsymbol{\sigma}_{xy}$.
\subsection{Oscillatory and stepwise shear for tissue linear response rheology}

To examine the mechanical behavior of the tissue, we impose oscillatory simple shear deformation on the simulated system using Lees–Edwards boundary conditions\cite{LE_pbc}. {Initial configurations were generated by first constructing a random point pattern, followed by Voronoi tessellation to obtain the initial cellular network. The resulting configuration was then relaxed according to Eqn.~\ref{EoM} with $\xi=0$ and $v_0=0$ using the FIRE algorithm~\cite{fire_algo_bitzek2006} until it reached a local energy minimum corresponding to a mechanically equilibrated, force-free metastable state. We refer to these configurations as strain-free because no external deformation is imposed during their preparation. However, they are not necessarily shear-stress-free, since residual internal stresses may remain after relaxation due to finite-size effects.} The affine deformation was subsequently applied according to the deformation gradient tensor $D=\left( \begin{smallmatrix} 1 & \gamma_0\sin(\omega t) \\ 0 & 1 \end{smallmatrix} \right)$. Following the affine deformation at every time step, the system was updated based on the overdamped dynamics described by Eqn.~\ref{EoM} with $v_0 = 0$ using Euler method. As we focus on the viscoelastic properties of tissues, T1 transitions were disabled during the oscillatory deformation process. 

To ensure that the system reaches steady state, after each deformation cycle $q$ of period $T=2\pi /\omega_0 $ for a particular driving frequency $\omega_0$, we compute the Fourier transform of the stress in the most recent 5 cycles:
\begin{equation}
    \Tilde{\sigma}_q(\omega_0)=\frac{1}{5T}\int_{(k-5)T}^{kT}\sigma(t)e^{i\omega_0 t}dt
\end{equation}
The system was considered to have reached steady state once $\Tilde{\sigma}_q(\omega_0)$ stabilized and no longer changed significantly with additional cycles, approaching a consistent value denoted by $\Tilde{\sigma}(\omega_0)$. The complex shear modulus was then calculated as $G^*(\omega_0) = \Tilde{\sigma}(\omega_0) / \Tilde{\gamma}(\omega_0)$, where $\Tilde{\gamma}(\omega_0)$ is the Fourier transform of the applied strain $\gamma(t)$.

To perform the numerical relaxation experiment, we applied a step strain to the system, holding the strain and letting the system relax. The corresponding deformation gradient tensor for this deformation is $D=\left( \begin{smallmatrix} 1 & \gamma_0\mathcal{H}(t) \\ 0 & 1 \end{smallmatrix} \right)$, where $\mathcal{H}(t)$ is the Heaviside function.

\subsection{Linear Rheological models}
\subsubsection{The Standard Linear Solid (SLS) model}

The Standard Linear Solid (SLS) model considered in this work consists of a spring with spring constant $E_0$ in parallel with a Maxwell component of elasticity $E_1$ and viscosity $\eta$. Since the material is composed of two parallel branches, the modulus is the sum of the moduli of the two branches, giving the relaxation modulus:
\begin{equation}
    G_{\text{SLS}}(t)=E_0+E_1e^{-tE_1/\eta}
    \label{sls G}
\end{equation}

and the complex modulus of the SLS material:
\begin{equation}
    G^*_{\text{SLS}}(\omega)=E_0+\frac{ i\omega E_1\eta }{E_1+i\omega \eta }
    \label{sls G star}
\end{equation}

\subsubsection{The Burgers model}
The Burgers model considered here consists of two Maxwell components ($E_1,\ \eta_1,\ E_2,\ \eta_2$) in parallel. The relaxation modulus of this Burgers model is:
\begin{equation}
    G_{\text{Burgers}}(t)=E_1e^{-tE_1/\eta_1}+E_2e^{-tE_2/\eta_2}
\end{equation}
and the complex modulus is:
\begin{equation}
    G_\text{Burgers}^*(\omega)=\frac{ i\omega E_1\eta_1 }{E_1+i\omega \eta_1} +\frac{ i\omega E_2\eta_2 }{E_2+i\omega \eta_2} 
    \label{burger G star}
\end{equation}
\subsubsection{The Fractional Burgers model}
Our Fractional Burgers model consists of a Maxwell component in parallel with a fractional Maxwell component. The fractional Maxwell component consists of a spring $E_1$ in series with a springpot with exponent $\beta$ and strength $c_\beta$. The creep compliance of the springpot is $J_{\beta}(t)=\frac{t^\beta}{c_\beta\Gamma(1+\beta)}$~\cite{frac_rheo_review}. Therefore, the creep compliance of the fractional Maxwell material is:
\begin{equation}
J(t) = \frac{1}{E_1}+\frac{t^\beta}{c_\beta\Gamma(1+\beta)}
\end{equation}
The Laplace transform of the creep compliance is:
\begin{equation}
J(s)=\frac{1}{E_1s}+\frac{1}{c_\beta s^{\beta+1}}
\end{equation}
Using the fundamental relation between relaxation modulus and creep compliance, $G(s)J(s)=1/s^2$~\cite{linear_viscoelas}, the Laplace transform of the relaxation modulus of the fractional Maxwell material can be obtained:
\begin{equation}
G_{\text{f-Maxwell}}(s)=\frac{1}{J(s)s^2}=\frac{E_1s^{\beta-1}}{s^\beta+E_1/c_\beta }
\end{equation}
The relaxation modulus can then be obtained by taking the inverse Laplace transform of $G(s)$, exploiting the integral representation of the Mittag–Leffler function~\cite{mittag_leffler_func}:
\begin{equation}
G_\text{f-Maxwell}(t)=E_1\mathcal{M}_{\beta,1}\big(-\frac{E_1}{c_\beta }t^\beta\big)
\label{frac maxwell G}
\end{equation}
where $\mathcal{M}_{a,b}$ is the Mittag–Leffler function with parameters $a$ and $b$. The relaxation modulus of a fractional Burgers model can then be obtained by adding the contribution of the conventional Maxwell branch to the modulus\cite{generalized_frac_rheo_solution}:
\begin{equation}
    G_\text{f-Burger}(t)=E_1\mathcal{M}_{\beta,1}\big(-\frac{E_1}{c_\beta }t^\beta\big)+E_2e^{-tE_2/\eta}
    \label{frac burger G}
\end{equation}

To obtain the complex modulus of the Fractional Maxwell model, we exploit the linear rheological assumption that under oscillatory strain $\gamma = e^{i\omega t}$, the stress is also harmonic $\sigma = G^*e^{i\omega t}$. This assumption can be combined with the constitutive equation $\gamma(t) = \frac{\sigma(t)}{E_1} +\frac{\sigma(t) t^\beta}{c_\beta\Gamma(1+\beta)}$ to obtain the complex modulus of the fractional Maxwell branch:
\begin{equation}
G^*_{\text{f-Maxwell}}(\omega)=\frac{E_1c_\beta(i\omega)^\beta}{E_1+c_\beta(i\omega)^\beta}
\label{frac maxwell G star}
\end{equation}
The complex modulus of the Burgers model therefore is\cite{generalized_frac_rheo_solution}:
\begin{equation}
    G^*_{\text{f-Burger}}(\omega)=\frac{E_1c_\beta(i\omega)^\beta}{E_1+c_\beta(i\omega)^\beta}+\frac{ i\omega E_2\eta }{E_2+i\omega \eta} 
    \label{frac burgers G star}
\end{equation}

\subsubsection{The Fractional SLS model}
\label{frac sls derivation}
In our Fractional SLS model, the conventional Maxwell branch is replaced by a fractional Maxwell component. Using Eqn.~\ref{frac maxwell G}, the relaxation modulus can be obtained as:
\begin{equation}
    G_{\text{f-SLS}}(t)=E_0+E_1\mathcal{M}_{\beta,1}\bigg(-\frac{E_1}{c_\beta}t^\beta\bigg)
    \label{frac sls G}
\end{equation}
Similarly, using Eqn.~\ref{frac maxwell G star}, the complex modulus of the fractional SLS is:
\begin{equation}
    G^*_{\text{f-SLS}}(\omega)=E_0+\frac{E_1c_\beta(i\omega)^\beta}{E_1+c_\beta(i\omega)^\beta}
    \label{frac sls G star}
\end{equation}

\subsection{Supplementary Figures}

\begin{figure}[htb]
    \centering
    \includegraphics[width=0.9\linewidth]{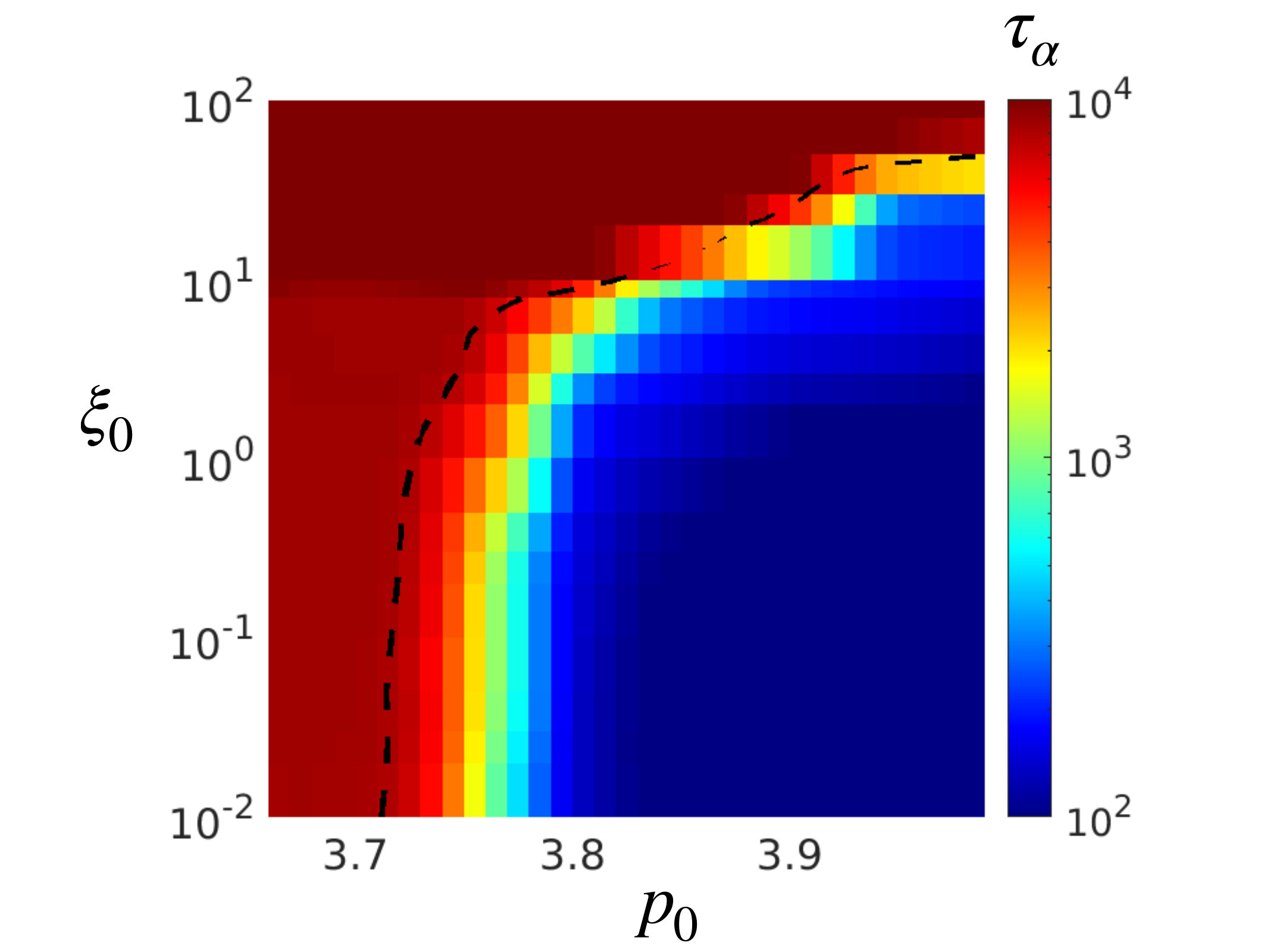}
    \caption{A second jamming phase diagram driven by cell--cell adhesion. Tissue states are quantified using the cell uncaging time $\tau_\alpha$, with the jamming–unjamming boundary defined by the contour $\tau_\alpha = 156$. This threshold is selected so that the phase boundary at $\xi_0 = 0$ aligns with the boundary obtained from the diffusivity-based criterion.}
    \label{tau alpha phase}
\end{figure}

\begin{figure*}
    \centering
    \includegraphics[width=1\linewidth]{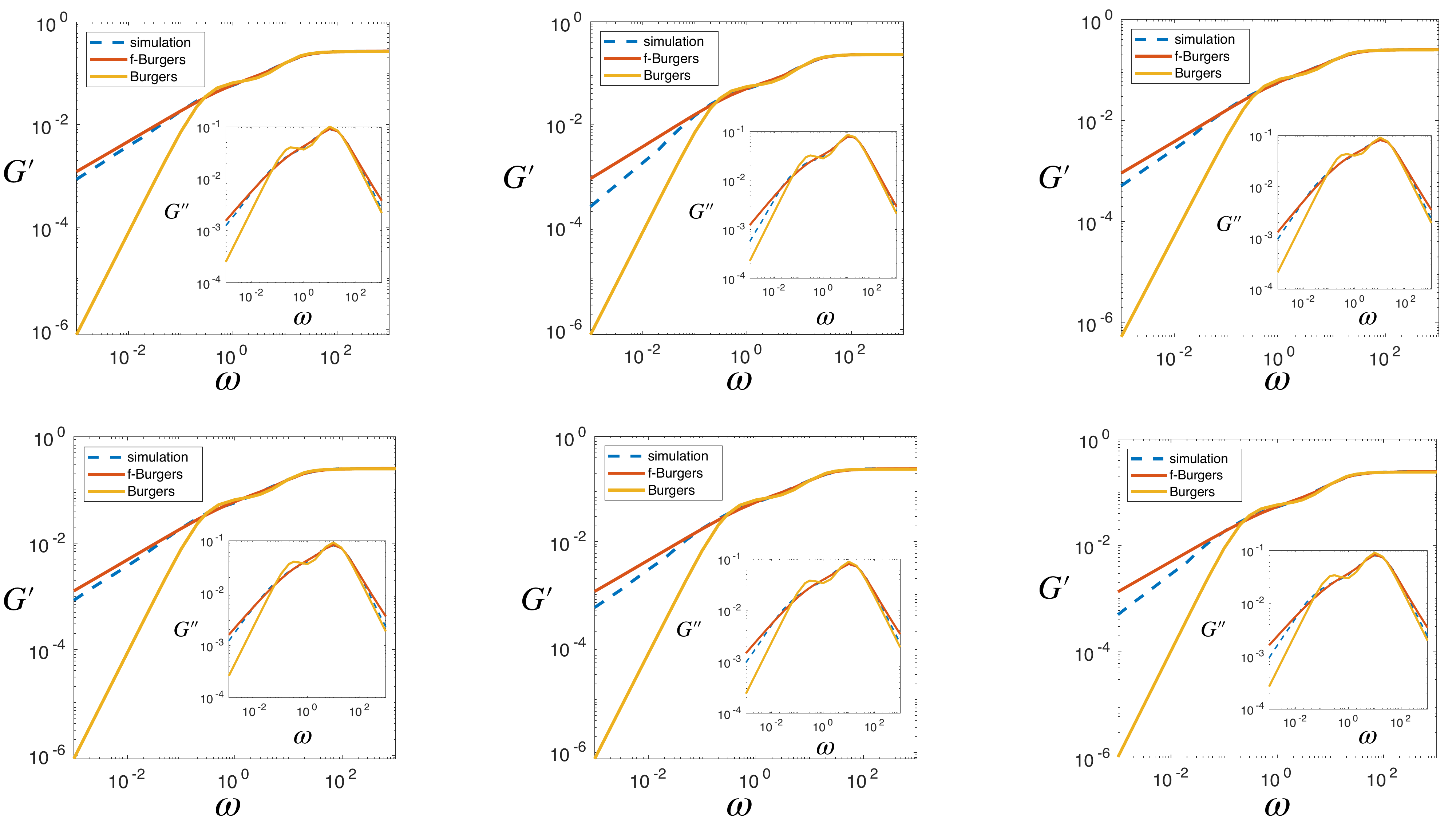}
    \caption{Complex moduli measured at $p_0 = 3.9$ and $\xi_0 = 0$ for multiple independent samples were fitted to both the classical Burgers model and the fractional Burgers model. The superior agreement with the fractional formulation demonstrates that the observed power-law rheology is intrinsic to the system and does not arise from sample-to-sample averaging.}
    \label{fluid complex mod}
\end{figure*}

\begin{figure*}
    \centering
    \includegraphics[width=1\linewidth]{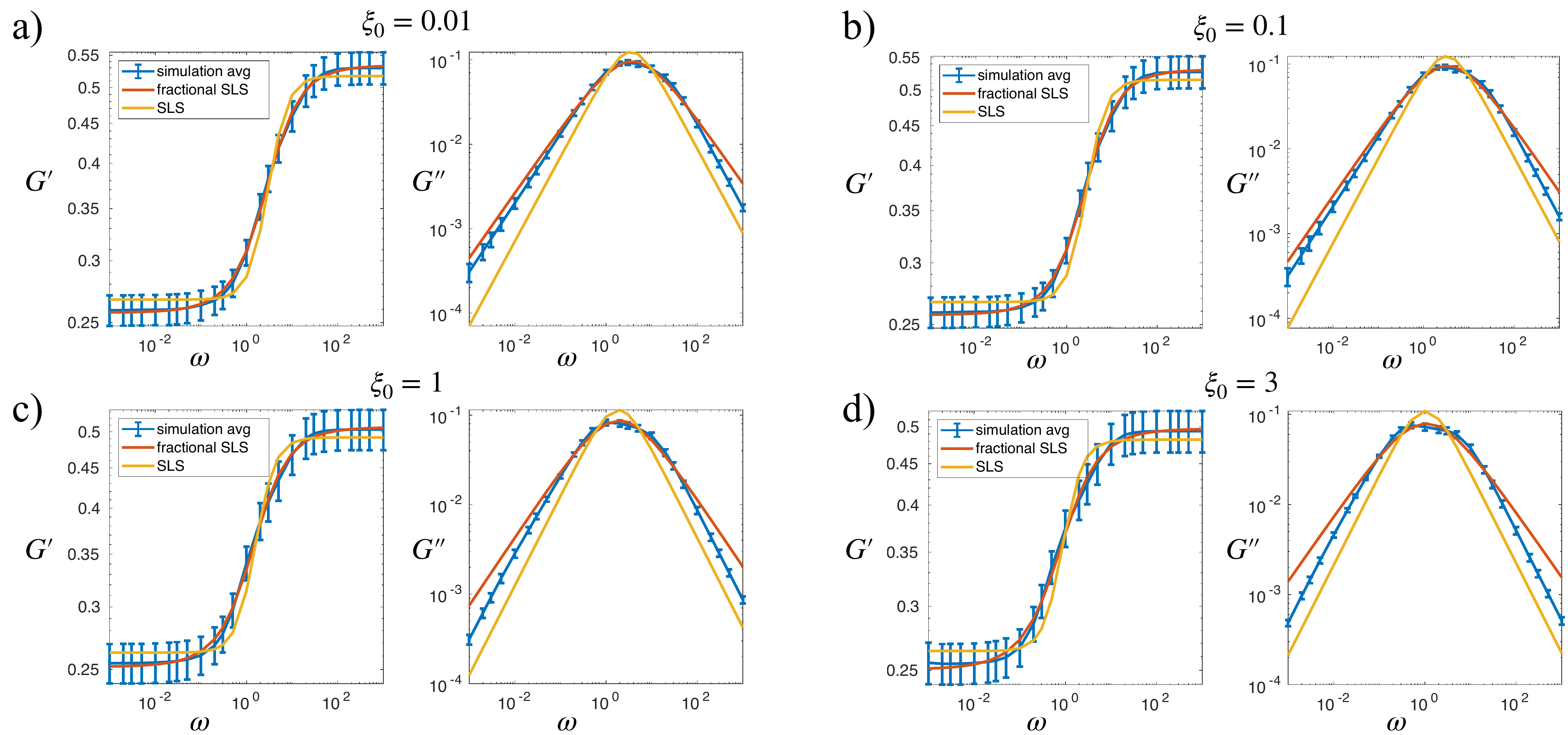}
    \caption{The storage and loss moduli at $p_0 = 3.72$ across different values of $\xi_0$, showing that the fractional SLS model consistently outperforms the conventional SLS model in capturing the frequency-dependent stress response.}
    \label{p3.72 complex mod}
\end{figure*}

\begin{figure*}[htb]
    \centering
    \includegraphics[width=0.4\linewidth]{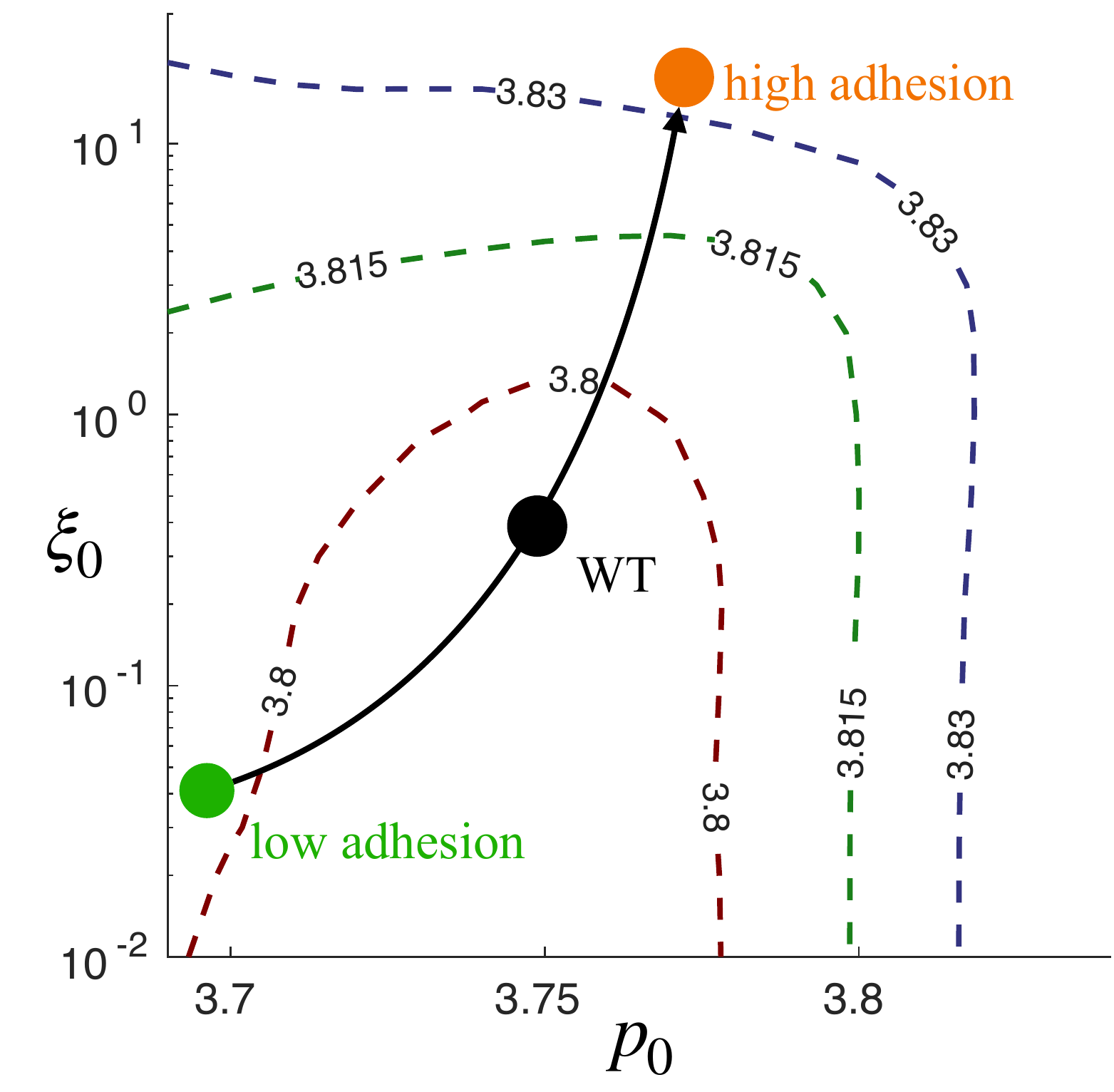}
    \caption{A hypothetical trajectory in the phase diagram illustrating the non-monotonic dependence of cell shape on adhesion strength. Along a path of increasing adhesion—represented here by a curve with positive slope—the cell shape may first decrease and subsequently increase, consistent with observations in the \textit{Drosophila} embryo.}
    \label{hypo q path}
\end{figure*}

\begin{figure*}[htb]
    \centering
    \includegraphics[width=0.8\linewidth]{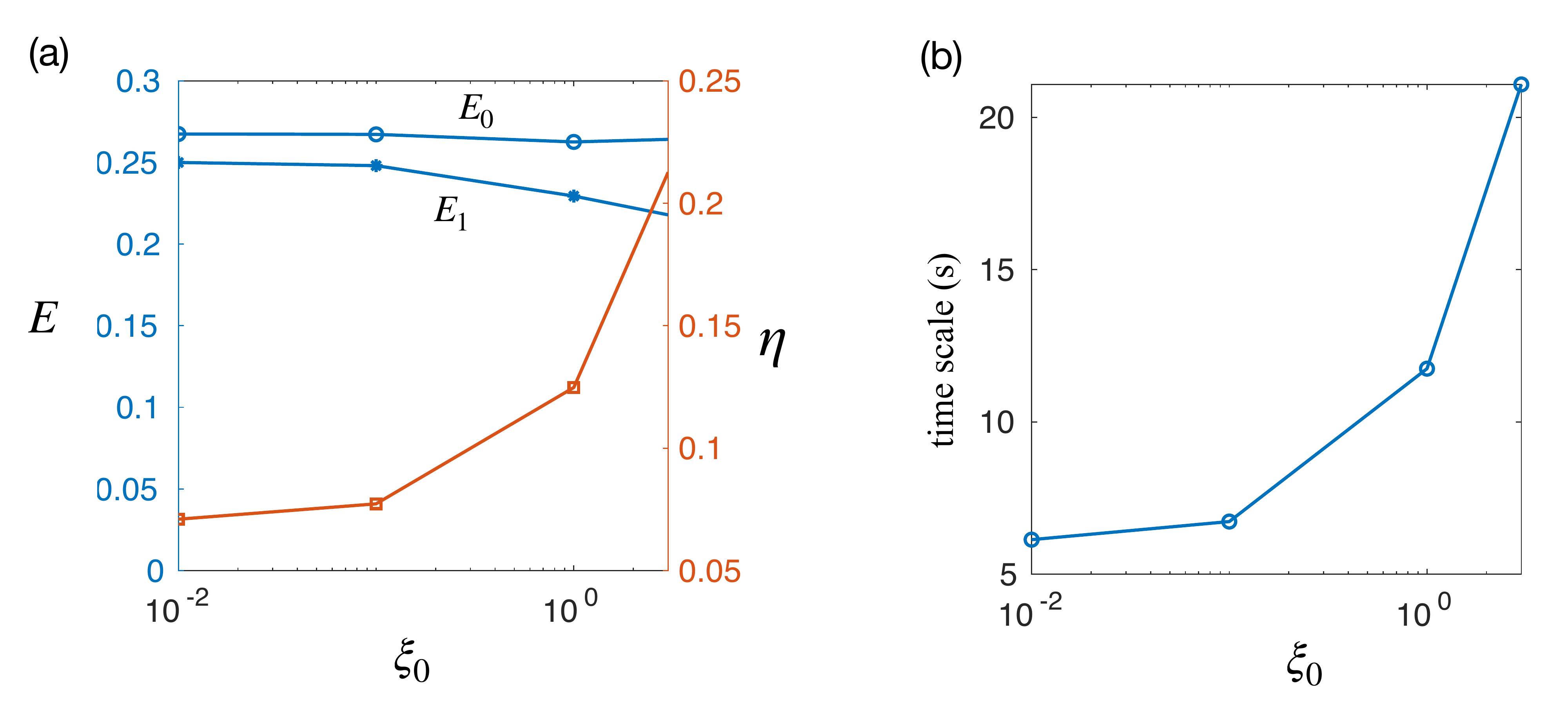}
    \caption{Extraction of the relaxation timescale from SLS model fits.(a) Standard Linear Solid (SLS) fitting parameters as a function of viscous adhesion $\xi_0$. (b) Using the time scale calibration reported in~\cite{bera2026shape}, the relaxation timescale $\tau = \eta/E_1$ is converted into physical time units, yielding values consistent with experimentally measured stress relaxation times in microtissues~\cite{walker2020time}.}
    \label{sls time scale}
\end{figure*}

\end{document}